\documentclass[journal]{IEEEtran}
\usepackage[export]{adjustbox}
\usepackage{amsmath}
\usepackage[usenames, dvipsnames, table]{xcolor}
\usepackage{multirow}
\usepackage{cite}
\usepackage[pass]{geometry}
\usepackage[ruled]{algorithm2e}
\usepackage{subfig}
\usepackage{array}

\usepackage{makeidx}         % allows index generation
\usepackage{graphicx}        % standard LaTeX graphics tool
\usepackage{multicol}        % used for the two-column index
\usepackage[bottom]{footmisc}% places footnotes at page bottom
\usepackage{amssymb,amsmath,amsthm,amsfonts}

 \usepackage[hyphens]{url}
\usepackage{breakurl}

\usepackage{dsfont}

\usepackage{todonotes}
\usepackage{color}

\usepackage{multirow}

\SetAlFnt{\small}
\SetAlCapFnt{\small}
\SetAlCapNameFnt{\small}
\SetAlCapHSkip{0pt}
\IncMargin{-\parindent}

\usepackage{xspace}

\mathchardef\mhyphen="2D

\renewcommand{\paragraph}[1]{\medskip \noindent \textbf{#1.\ }}

\usepackage{mathtools, nccmath}
\usepackage{color, colortbl}
\usepackage{tabularx}
\usepackage{enumitem}
\usepackage{rotating}
\usepackage{booktabs}
\usepackage{epstopdf}

\usepackage{tikz}
\let\svtikzpicture\tikzpicture
\def\tikzpicture{\noindent\svtikzpicture}
\usetikzlibrary{shapes,arrows}
\usetikzlibrary{positioning}
\usetikzlibrary{shadows}
\usepackage{relsize}

\makeatletter
\newcommand{\mypm}{\mathbin{\mathpalette\@mypm\relax}}
\newcommand{\@mypm}[2]{\ooalign{%
  \raisebox{.1\height}{$#1+$}\cr
  \smash{\raisebox{-.6\height}{$#1-$}}\cr}}
\makeatother

\title{Source Camera Attribution of Multi-Format Devices}
\author{Samet Taspinar,
        Manoranjan Mohanty,
        and Nasir Memon
\thanks{Samet Taspinar (st89@nyu.edu) is with Center for Cyber Security, New York University Abu Dhabi, UAE, Manoranjan Mohanty (m.mohanty@auckland.ac.nz) is with Department of Computer Science, University of Auckland, and Nasir Memon (memon@nyu.edu) is with Department of Computer Science and Engineering, New York University, New York, USA.}
}

\begin{document}
\maketitle
\begin{abstract}
 
Photo Response Non-Uniformity (PRNU) based source camera attribution is an effective method to determine the origin camera of visual media (an image or a video). However, given that modern devices, especially smartphones, capture images, and videos at different resolutions using the same sensor array, PRNU attribution can become ineffective as the camera fingerprint and query visual media can be misaligned. We examine different resizing techniques such as binning, line-skipping, cropping and scaling that cameras use to downsize the raw sensor image to different media. 
Taking such techniques into account, this paper studies the problem of source camera attribution. We define the notion of Ratio of Alignment, which is a measure of \textit{shared sensor elements} among spatially corresponding pixels within two media objects resized with different techniques. We then compute the Ratio of Alignment between the different combinations of three common resizing methods under simplified conditions and experimentally validate our analysis. 
Based on the insights drawn from the different techniques used by cameras and the RoA analysis, the paper proposes an algorithm for matching the source of a video with an image and vice versa. We also present an efficient search method resulting in significantly improved performance in matching as well as computation time.

\end{abstract}


\vspace{-4mm}
\section{Introduction}
The emergence of "fake news'' along with sophisticated techniques using machine learning to create realistic looking content such as deep-fakes has led to an increased interest in digital media forensics~\cite{TahaBook2013, fridrich2009digital, delp2009digital, milani2012overview}. One well-studied problem in digital media forensics is to discover the source of an image or a video. Photo-Response Non-Uniformity (\textit{PRNU}) based source camera attribution\cite{lukas2006digital} is a well-known technique that can determine whether a particular device was used to capture a specific visual object. Here, a PRNU camera fingerprint (or more precisely a fingerprint estimate) is first computed from multiple still images (i.e., images or video frames) known to be taken by a specific camera. Then, the PRNU noise extracted from a query visual media is correlated with this fingerprint to determine if it was taken with the given camera. 

To perform PRNU-based source camera attribution, the query visual media has to be precisely aligned with the camera fingerprint. That is, the $(i,j)^{\rm th}$ pixels of the fingerprint and query images should correspond to largely the same elements of the camera sensor array. When misalignment between the fingerprint and query image occurs due to simple geometric transformations such as resizing and cropping, attribution can still be made by exhaustively trying all the possible transformation parameters \cite{jessica:crop}. However, this can be a very time-consuming process. Efforts at speeding this up have been proposed achieving a speed-up factor of around ten by downsizing the media to be matched \cite{yaqub2018towards}.

Although simple misalignment can be compensated for by exhaustive search techniques, some recent anonymization methods to prevent source camera attribution create complex misalignment using techniques like seam carving that make exhaustive search intractable ~\cite{bayram2013seam, dirik2014analysis}. In the case of seam carving, it was subsequently shown that when multiple seam carved images are available from the same camera, successful verification could still be done by increasing alignment between the camera fingerprint and the seam-carved images~\cite{taspinar2017prnu, taspinar2016prnu}, provided no additional operation such as scaling and cropping has been performed. In-painting \cite{mandelli2017inpainting}, Patch-based desynchronization~\cite{John:2016:Patch}, and image stitching~\cite{Karak:2015:PRNU} are other examples of complex techniques for breaking alignment.

Complex misalignment between a camera fingerprint and query object also occurs when they represent different types of media. For example, this happens when the camera fingerprint has been computed from images and the query object is a video captured at a different resolution. Given that modern devices such as smartphones can capture different types of media with different resolutions, and given that social networks often transform visual media objects in different ways, performing source camera attribution with different types of media, potentially from different social platforms, and taken from the same camera is a real and relevant problem. Recently, DARPA's Medifor program \cite{darpa2018medifor} issued a challenge for camera identification with a dataset that included \textit{same-type media} (i.e., image to image or video to video) as well as \textit{mixed media} (i.e., matching images to videos or vice versa). 

In this paper, we study the problem of camera verification in the context of mixed media. The attribution scenarios we examine include \textit{a video vs a single image, a video vs a set of images}. For each of these cases, one of the visual objects is from a known source, and the other is a query object whose source is under question. 

The main contributions of this paper are summarized below. 
\begin{itemize}
  \item We undertake a comprehensive study of source camera attribution with mixed media, taking various factors into account such as different aspect ratios, different techniques for capturing low-resolution content and different parts of the sensor used for media capture. These factors have not been taken into account by media forensics researchers before and partly explain the state-of-the-art results presented in the paper.
  \item We define the notion of {\em alignment} and {\em Ratio of Alignment (RoA)} between two still images taken from the same camera but using different capture techniques. The notion of RoA provides a simple and intuitive framework to better explain and quantify why PRNU based attribution may or may not work when there is a mismatch between resizing techniques used in practice and in testing. 
  \item We provide an analytic determination of the ratio of alignment (albeit for a simplified case) between the three most common resizing techniques, namely, binning, line-skipping and bilinear scaling.

  \item Based on our analysis and experimental validation, we explore why and how different resizing techniques, namely, bilinear scaling, line-skipping and binning perform for source camera matching with mixed media. We propose an algorithm that represents a good balance between performance and computing when matching a video and an image (and vice versa). 

  \item Given the importance of performing attribution in the presence of scaling and cropping, we propose an efficient search method that is significantly faster than naive exhaustive search.
  \item We compile experimental results that lead to insights on parameters used by numerous camera brands and models with respect to the in-camera operations they use for capturing image and video content. This knowledge can be used to significantly speed up attribution when the camera model of the object is known. 
  \item We compile a mixed media dataset to be shared with the community, that contains images and videos of multiple resolutions from a variety of different cameras.
\end{itemize}

The rest of this paper is organized as follows. Section~\ref{sec:background} provides an overview of PRNU-based source camera attribution and describes different approaches used for capturing images and video frames. Section~\ref{sec:math} defines the notion of `Ratio of Alignment" and provides its analytic determination between three common techniques, namely, bilinear scaling, binning and line-skipping. In Section~\ref{sec:approach}, we examine camera attribution for different scenarios of mixed media formats that a forensic analyst may encounter. In Section \ref{sec:experiment} we provide experimental results. Section~\ref{sec:conclusion} concludes the paper.

\vspace{-3mm}
\section{Background}
\label{sec:background}
\vspace{-1mm}
In this section, an overview of how a camera captures a video and an image using a single sensor array is first provided. Then, we describe different resizing techniques that modern cameras use when capturing a video. Finally, we briefly summarize PRNU-based camera attribution.

% \vspace{-3mm}
% \subsection{In-camera processing}
% \label{section:background:incamera}
% This subsection first provides a very high-level overview of general imaging pipeline. It then describes in-camera resizing techniques that are used to capture a media object that has a lower resolution than the camera sensor. Finally, the notion of active image boundary, a process that can have an impact on mixed-media attribution, is explained.

\vspace{-3mm}
\subsection{Image capturing pipeline} 
\begin{figure}[!ht]
\centering
  \includegraphics[width=75mm,trim={2.5cm 15.6cm 9.9cm 1.9cm},clip, right]{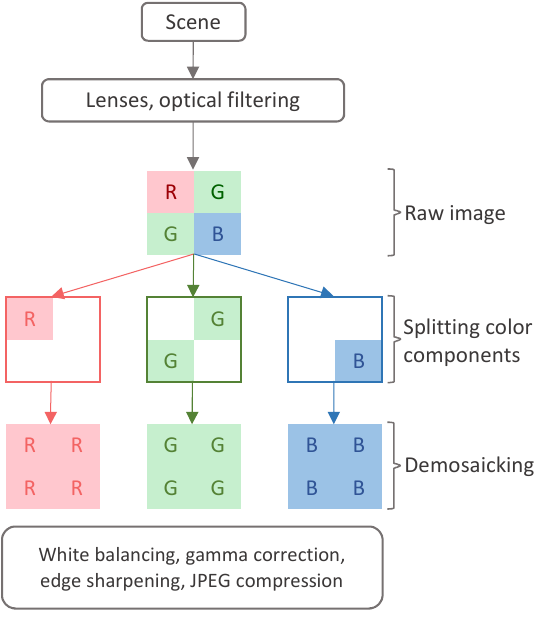}
  \caption{{\color{black}Imaging pipeline of a digital camera}}
  \label{Fig:approach:pipeline}
\end{figure}

There is much processing that takes place within a camera after light from a scene is guided through the lens to the sensor array. Although different camera models may apply different processing steps, many of these steps are common to most cameras. Fig.~\ref{Fig:approach:pipeline} shows a simplified imaging pipeline.

Most single-chip cameras use a color filter array (CFA) which arranges RGB filters on a square grid. Hence, each sensor element receives only one of red, green, or blue components of the light passing through the lens. There are many different patterns according to which a CFA can be configured. The most common pattern is known as the Bayer Filter \cite{bayer1976color} and one particular variation of it is shown in Fig.~\ref{Fig:approach:pipeline} where every $2 \times 2$ pixel array comprises of two green pixels and one red and one blue pixel. The missing color values in each pixel are interpolated from corresponding color values of neighboring pixels to get a full-color image. 

After demosaicing, the remaining steps are mainly related to image quality. These include, for example, white balancing, gamma correction, and edge sharpening. Finally, JPEG compression is applied to the image which significantly decreases the disk storage needed with a negligible perceptible loss in quality. It should be noted that none of these steps cause any geometrical transformation of the image. 

\vspace{-3mm}
\subsection{In-camera resizing}
\label{background:sensor_resizing}

Modern cameras typically contain over 10 million pixels which help capture intricate scene details in an image. However, as the number of pixels increase, the computational cost to capture a still image also increases. Thus, most cameras don't use the full sensor resolution when capturing a video and downsize the sensor output to a lower resolution by in-camera processing. Moreover, based on user settings, images can also be captured at a lower resolution.
To downsize an image or video frame when the sensor resolution is higher than the desired resolution, a combination of cropping, line-skipping, binning, and some scaling methods can be applied to the media. We describe these techniques below. 

\vspace{-2mm}
\begin{figure}[!ht]
  \centering
  \includegraphics[trim={2.5cm 22.4cm 8.3cm 4.0cm},clip,width=87mm] {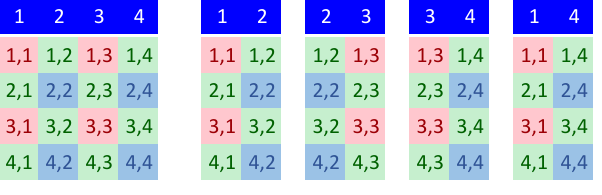}
  \vspace{-2mm}
  \caption{{\color{black}Four ways of resizing using line-skipping }}
  \label{Fig:ls_ex}
\end{figure}

\begin{figure}[!ht]
  \centering
  \subfloat[An example of line-skipping]
  {\label{Fig:lineskip}\includegraphics[width=60mm]{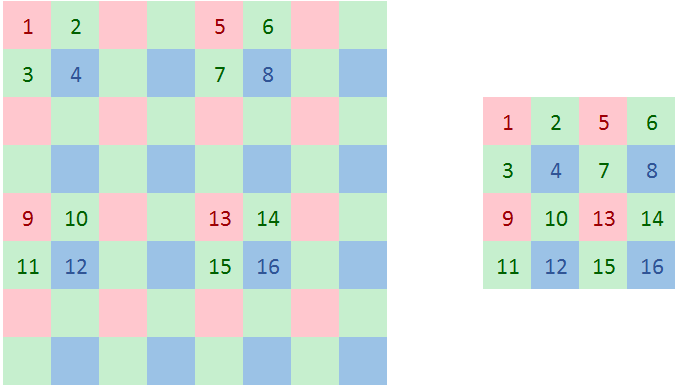}}

  \subfloat[$2\times 2$ binning scheme (Obtained from ~\cite{jin2012analysis})]
  {\label{Fig:binning}\includegraphics[width=60mm]{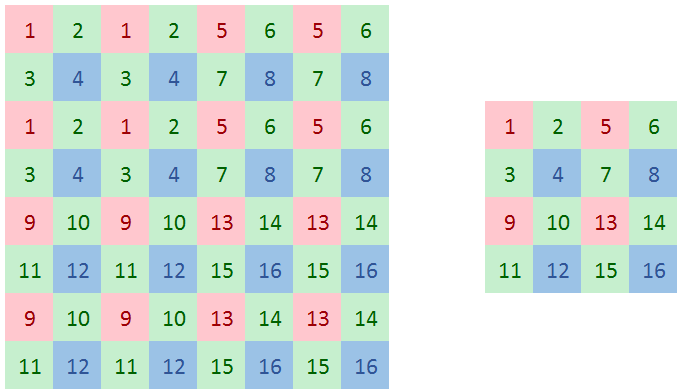}}
  \caption{Two different in-camera resizing schemes}
  \label{Fig:resizing schemes}
% \vspace{-2mm}
\end{figure}

%\vspace{-1mm}
\paragraph{Line-skipping} Line-skipping is a technique that omits all the pixels in a row and/or column. {\color{black} After omitting the lines, the output image is expected to maintain a valid Bayer pattern. This can be done in several different ways. Fig.~\ref{Fig:ls_ex} shows an example where a $4\times 4$ image is resized to $4\times 2$ by skipping $2$ of the $4$ columns (i.e., $1,2,3,4$). This can be achieved with one of four possible ways with a valid Bayer pattern as shown in Fig.~\ref{Fig:ls_ex}.
If we extend this to both axes to downsize the image to $2\times 2$ resolution, we can now have $16$ different ways. For example, in Fig.~\ref{Fig:lineskip}, pairs of columns and rows are alternately kept and skipped, starting with keeping the $1^{\rm st}$ and $2^{\rm nd}$. }

\vspace{-1mm}
\paragraph{Pixel binning} Pixel-binning combines the values of multiple pixels of the same color in the raw image to create a composite pixel. For example, {\color{black}in Fig.~\ref{Fig:binning}, four red pixels on the left (i.e., the groups labeled $1, 5, 9,$ and $13$) are combined together to create the red pixels in the binned image shown on the right. The green and blue pixels are also are created in the same way. We illustrate $2\times 2$ pixel binning, however, it is possible to do $k\times k$ binning which downsizes an image by $1/k$} 

{\color{black}To the best of our knowledge, weighted pixel-binning for resizing images to an arbitrary resolution is not used by cameras as this is against one of the main goals achieved by binning, i.e., decreasing computation of video capture. If cameras choose to further downsize still images after binning, they use another scaling technique, such as bilinear scaling. Binning can be enabled in a camera only if it is needed \cite{zhang2018pixel}. 
}

\vspace{-1mm}
\paragraph{Scaling} {\color{black} As we mentioned, users can choose to capture images with a lower resolution. So resizing is not limited to video capture. Along with the techniques above, cameras use other scaling techniques such as bicubic or Lanczos (or their derivatives) to downsize still images (i.e., images or video frames). 
%This can be achieved with one of these methods that we mentioned. 
Some cameras with high computational power may not use binning or line-skipping for videos; they can simply process the entire sensor data and downsize using a scaling technique at the end of imaging pipeline.}
% https://web.stanford.edu/group/vista/cgi-bin/wiki/index.php/Pixel_Binning#Introduction

\begin{figure}[!ht]
 \centering
 {\color{black}
 \subfloat[Downsizing by half by cropping from the center.]
 {\label{Fig:l1}\includegraphics[height=30mm]{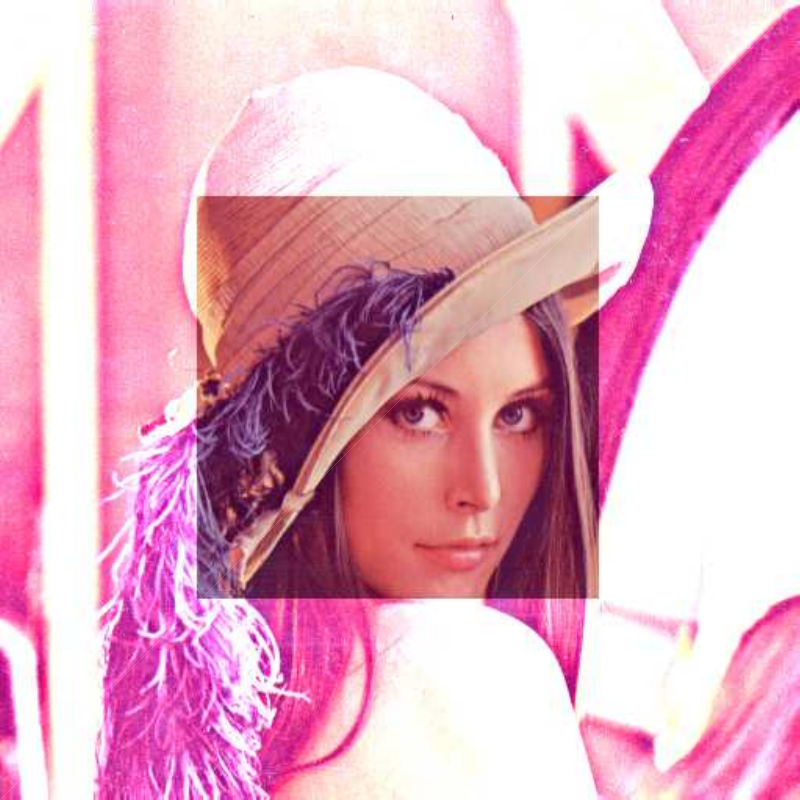}}~~~~~
 \subfloat[Downsizing by $5/6$ by cropping from the center. Then resized by $3/5$ to obtain the right image]
 {\label{Fig:l2}\includegraphics[height=30mm]{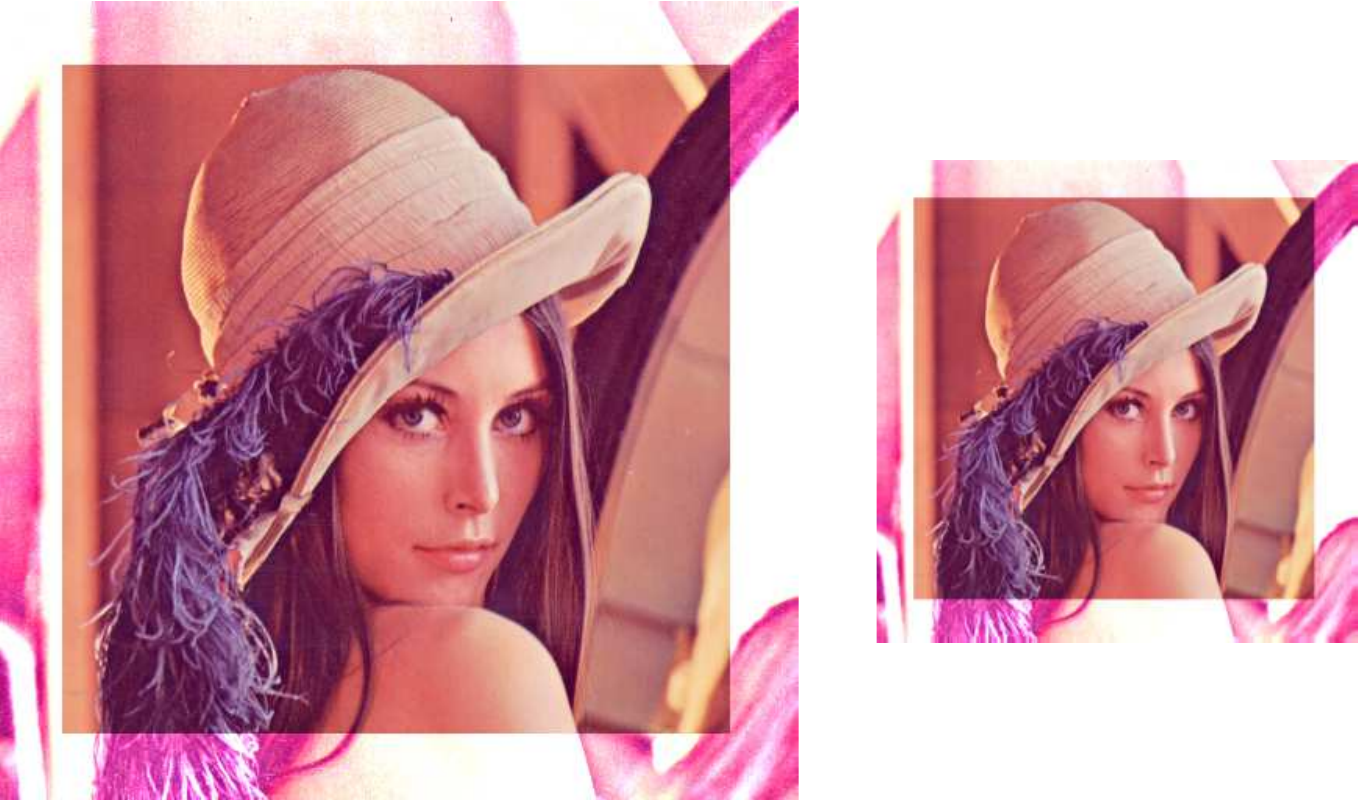}}
 \caption{{\color{black}The effect of cropping on field of view: The sensor captures the full resolution image. The saturated regions (borders) are discarded and the centers are the final images.}}
 \label{fig:field_view}}
\end{figure}

\paragraph{Cropping and Scaling} {\color{black} Cropping is the simplest technique to decrease image resolution. One approach is to use only the central pixels of a sensor and discard the surrounding pixels of the region of interest.
However, one of the biggest drawbacks of cropping is that it changes the field of view in a still image if the cropped area is large. Hence it is cropping is most often used along with resizing. 
% Here, we compare an image that is resized to half resolution using only cropping and obtaining the same resolution with cropping along with resizing. 
Fig.~\ref{fig:field_view} shows two examples where the original image is downsized to half resolution. In Fig.~\ref{Fig:l1}, the image is cropped from the center without any resizing operation. Whereas in Fig.~\ref{Fig:l2} downsizing by $5/6$ is first done by cropping from the center. Then resizing by $3/5$ is performed to make it half resolution. As seen in the final outputs (the unsaturated parts of Fig.~\ref{Fig:l1} and Fig.~\ref{Fig:l2} [rightmost image]), both images are of the same resolution, but a significantly wider region of the scene is captured when using the second approach. Note that the resizing step can use any of the three techniques described above.}

% \vspace{-2mm}
\begin{figure}[!ht]
  \centering
  \includegraphics[trim={4.1cm 13cm 11.4cm 10.4cm},clip,width=42mm] {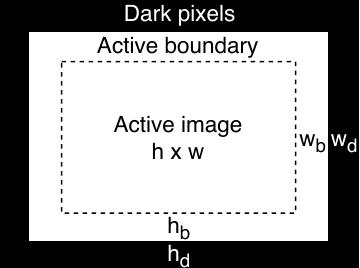}
  \caption{{\color{black}Different regions of camera sensor}}
  \label{Fig:boundary}
\end{figure}
\vspace{-1mm}
\paragraph{Active image and active boundary} In some cameras (such as Nexus 6, Lenovo P1~\cite{taspinar2016source}), the size of the sensor array is larger than listed in the camera specification as shown in Fig.~\ref{Fig:boundary}. Here, the \textit{active image}, the center of the sensor as per the resolution in the specification, is surrounded by an \textit{active boundary} of pixels. The active boundary is further surrounded by \textit{dark pixels}, which are used for monitoring the black color level. The active image region of the sensor is typically used for capturing an image whereas both the active image and the active boundary regions are used when capturing a video \cite{imaging2004mt9m001}. Therefore, a video may contain some of the boundary pixels which were not used while capturing an image. It is crucial to take boundary pixels into account while correlating a video and an image; otherwise, they may fail to match even if they are from the same source camera.

\vspace{-2mm}
\subsection{PRNU-based source camera attribution}
\label{sec:background:PRNU}
PRNU-based camera attribution is established on the fact that the output of the camera sensor, $I$, can be modeled as 
\vspace{-1mm}
\begin{equation}
  I = I^{(0)} + I^{(0)}X + \psi
\label{Eqn:im_noise}
\end{equation}
% \vspace{-2mm}
where $I^{(0)}$ is the noise-free still image, $X$ is the PRNU noise, and $\psi$ is a combination of additional noise, such as readout noise, dark current, shot noise, content-related noise, and quantization noise. Denoising is typically done in each color component separately which results in three PRNU noise patterns of the same resolution: $X_R$, $X_G$, and $X_B$ \cite{fridrich2009digital}. These three noise components are then converted to a final noise component $X$ as follows:
\vspace{-1mm}
\begin{equation}
\label{eqn:noise_weights}
  X=0.3\times X_R + 0.6\times X_G + 0.1\times X_B.
\end{equation}
\vspace{-2mm}

Since denoising filters (such as Wavelet Denoising\cite{mihcak1999low}, BM3D\cite{dabov2009bm3d}) are not perfect, they cannot totally eliminate the random noise, $\psi$. Hence, multiple still images are averaged to minimize $\psi$ and improve the estimation of $X$, which is called the camera fingerprint. 
A given query image can then be attributed to a camera by matching the PRNU noise extracted from the query image with $X$ using the Pearson correlation coefficient or Peak-to-Correlation Energy (PCE). However, for this to work, the PRNU of the query image has to be aligned with the camera fingerprint. If the image or fingerprint is resized, the correct resizing parameter must be found, and the resizing operation must be reversed. A brute force method can be used to find the resizing parameters~\cite{jessica:crop}. When a still image is cropped, Normalized Cross Correlation (NCC)~\cite{lewis1995fast} can be used to find the cropping location~\cite{jessica:crop}.

Although PRNU-based source camera attribution has been well studied, enough attention has not been given towards attribution in the presence of mixed media datasets that contain both videos and images. So far, most of the work has focused on either images or videos (but not both).
For images, much work has been done to improve PRNU based attribution~\cite{jessica:camera, Sutcu:2007:Improve, Li:2012:Color, Chierchia:2010:IDP, Li:2010:Enhance},
as well as using the scheme for purposes other than attribution~\cite{Caldelli:201:WIFS, lukavs2006detecting, Chierchia:2014:Fing}. Many researchers have also extended image-centric methods towards video ~\cite{Simone:VideoOv:2012, Chen:2007:Vid, McCloskey:2008:Confidence, Chuang:2011:Vcomp, Chen:2013:Video, Houten:2009:YVid, hyun2012camcorder, taspinar2016source}. Taspinar et al. \cite{taspinar2016source} and Iuliani et al. \cite{iuliani2017hybrid} addressed attribution of multi-format devices with limited success using a brute-force search.

\vspace{-3mm}
\section{Matching in Presence of Desynchronization}
\label{sec:math}
\medmuskip=1mu
\thinmuskip=1mu
\thickmuskip=1mu

\newcommand{\tmin}{\text{-}}

As we have seen in the previous section, different methods can be employed for capturing different types of media with different resolutions. A lower resolution media/fingerprint could be matched with a higher resolution media/fingerprint if the exact techniques and parameters that were used to create the lower resolution media are known. But this is generally not feasible as device manufacturers do not reveal such information. Besides in many situations, such as media gathered from social networking sites, there may be no information available about the camera model at all. 

%In this section, we define the notion of alignment and ratio of alignment between two media captured using the same sensor. We provide an analytic determination of the ratio of alignment for two specific cases. Specifically, alignment between an image downscaled using bilinear scaling and a non-stabilized video frame, acquired at half the resolution of the image, using binning and line-skipping respectively. We show that source camera attribution is still possible by matching these objects as the ratio of alignment is significantly high. We also show experimentally that better attribution results with a higher ratio of alignment. 

{\color{black} However, as is known in the media forensics research community, attribution may still be possible. For example,} As established earlier with seam carving ~\cite{taspinar2017prnu, taspinar2016prnu},  even if an image has been downsized in a complex manner and is no longer synchronized with a camera fingerprint, attribution may still be possible with a suitably  scaled down fingerprint if there are common pixels from the original versions that have been used to compute each pixel in the downsized version. The same idea applies when matching different media taken from the same sensor.  As an example, consider Fig.~\ref{Fig:resizing schemes}. It can be seen that the set of input pixels used for computing a particular output pixel using binning and line-skipping can have common pixels. For example, for computing the pixel location $(1, 1)$ in the output image, binning uses input pixels $\{(1, 1), (1, 3), (3, 1), (3, 3)\}$ from the raw image, and line-skipping uses $\{(1,1)\}$.
In these sets, the input pixel $(1, 1)$ is common between binning and line-skipping. However, the contribution of a common pixel can differ depending on the resizing parameters. In this example, the contribution of $(1, 1)$ pixel is $25\%$ for binning and $100\%$ for line-skipping.  Note that this is a simplified example where the demosaicing step is neglected for both images. 

The presence of at least one common pixel in the input sets (such as $(1, 1)$ in the above example) may lead to successful PRNU-based correlation between two media objects\cite{taspinar2017prnu}. The more the pixels in common, the higher the correlation tends to be. To compare the degree and number of common pixels between two media objects taken from the same sensor but using different resizing strategies, we define the notion of \textit{Ratio of Alignment (RoA)} in the subsections below.  {\color{black} We then derive RoA for a few cases of different pairs of misaligned media arising from some common resizing approaches. Although the cases are simplified, the results provide insights that help formulate and understand better techniques for PRNU attribution.} First we provide some notation below. 

\vspace{-4mm}
\subsection{Notation for Ratio of Alignment (RoA) derivations} \label{sec:Syntax}
\begin{itemize}
    \item $I^{raw}$: A matrix representing raw sensor output before demosaicing. $I^{raw}$ has a resolution of $M \times N$. 
    \item For brevity, and where the context is clear, we use $I(m, n)$ to denote the value of $I^{raw}(m, n)$. The red, green, and blue components are at location $I(2m-1,2n-1)$,  $I(2m-1,2n)$ and $I(2m, 2n-1)$, and $I(2m,2n)$, respectively, for any $1 \leq m \leq M/2$ and $1 \leq n \leq N/2$.
    \item $I_C^{Bscale}(m, n)$, $I_C^{Bin}(m, n)$, and $I_C^{Line}(m, n)$ denote the $(m, n)^{\rm th}$ pixel of color component $C$ after $I^{raw}$ is resized by bilinear interpolation, binning, and line-skipping, respectively. The color component $C$ is one of red, green, or blue.
    \item $A$ is the RoA between two images and for $RGB$ images, it consists of three separate components, $A_R$, $A_G$, and $A_B$ for red, green and blue color planes, respectively.
    \item $I(initVal:step:endVal)$: Represents the formula $$\sum_{i=0}^{\lfloor (endVal-initVal)/step \rfloor} I(i \times step +initVal)$$ where $\lfloor~~\rfloor$ is floor operation. Note that this equation indicates the sum of the pixels from $initVal$ to $endVal$ with an increment of $step$. For example, $I(3:2:9) = I(3) + I(5) + I(7) + I(9)$. Also when the $step$ is skipped (\textit{i.e.} $initVal:endVal$), the value of $step$ is by default considered as one (\textit{i.e.} $initVal:1:endVal$). 
    \item $I(initX:stepX:endX, initY:stepY:endY)$: Indicates the sum of all pixels from $I(initX)$ to $I(endX)$ and $I(initY)$ to $I(endY)$ with an increment of $stepX$, and $stepY$ along $X$ and $Y$ axes, respectively. For example, $I(2:2:4,3:3:6) = I(2,3)+I(2,6)+I(4,3)+I(4,6)$. 
    \item The analysis is done for different cases based on whether pixel positions $m$ and $n$ are odd or even:
    \begin{itemize}
        \item case 1: $m, n$ are odd
        \item case 2: $m$ is odd, $n$ is even
        \item case 3: $m$ is even, $n$ is odd
        \item case 4: $m, n$ are even
    \end{itemize}
\end{itemize}

\vspace{-4mm}
\subsection{Ratio of Alignment (RoA) definition and example}
\label{sec:Ratio}
Before defining Ratio or Alignment (RoA), first, we define what we mean by alignment. Suppose the raw image, $I^{raw}$, is resized to half using two different resizing techniques (e.g., one is resized by binning and the other with bilinear scaling). Denote the first resized image as $I^{\prime} = \{I^{\prime}(1, 1), I^{\prime}(1, 2), \dots, I^{\prime}(M/2, N/2) \}$ and the second as $I^{\prime\prime} = \{I^{\prime\prime}(1, 1), I^{\prime\prime}(1, 2), \dots, I^{\prime\prime}(M/2, N/2) \}$. To evaluate the RoA between $I^\prime$ and $I^{\prime\prime}$, we first determine the alignment of $(i, j)^{\rm th}$ pixel. If there is at least one common pixel in the computation of $I^{\prime}(i,j)$ and $I^{\prime\prime}(i,j)$, then the two pixels are \textit{partially aligned}. If the two are computed from identical sets, then they are \textit{fully-aligned}. If all the pixels in $I^{\prime}$ are fully-aligned with their corresponding pixels in $I^{\prime\prime}$, then $I^{\prime}$ and $I^{\prime\prime}$ are said to be fully-aligned (i.e., both $I^{\prime}$ and $I^{\prime\prime}$ down-sized by the same resizing technique). For brevity, we say that two images are aligned, when the images are either fully or partially aligned, with the context making the specific case clear. 

Suppose the $(i,j)^{\rm th}$ pixel of the color component $C$ of $I^{\prime}$, $I^{\prime}_C(i,j)$, is computed from $n$ pixels from $I^{raw}$, $\{I^{raw}(p_1),I^{raw}(p_2),...,I^{raw}(p_n)\}$ with weights $\{w^{\prime}(p_1),w^{\prime}(p_2),...,w^{\prime}(p_n)\}$ (here $p_1,\dots p_n$ are pixel indices in a vector form), whereas the pixel $I^{\prime\prime}_C(i,j)$, is computed from the $m$ pixels $\{I^{raw}(q_1),I^{raw}(q_2),...,I^{raw}(q_m)\}$ with weights $\{w^{\prime\prime}(q_1),w^{\prime\prime}(q_2),...,w^{\prime\prime}(q_m)\}$. Now suppose their intersection consists of $l$ pixels, $\{I^{raw}(k_1), I^{raw}(k_2), ..., I^{raw}(k_l)\}$. 
Then, the alignment between $I^{\prime}_C(i,j)$ and $I^{\prime\prime}_C(i,j)$ is defined as
\vspace{-1mm}
\begin{equation}
    a_C\big(I^\prime(i,j),I^{\prime\prime}(i,j)\big)  = \sum_{u=1}^{l} min\big(w^{\prime}(k_u), w^{\prime\prime}(k_u)\big)
    \label{eqn:alignment4px_new}
\end{equation}

To obtain the alignment between $I^{\prime}$ and $I^{\prime\prime}$ over all pixels in the color plane $C$, $A_C(I^\prime, I^{\prime\prime})$, we average the alignment, $a_C\big(I^\prime(i,j),I^{\prime\prime}(i,j)\big)$, over all the pixels as follows:

\vspace{-1mm}
\begin{equation}
    A_C(I^\prime, I^{\prime\prime}) = \frac{1}{M/2 \times N/2} \sum_{i=1}^{M/2} \sum_{j=1}^{N/2} a_C\big(I^\prime(i,j),I^{\prime\prime}(i,j)\big) 
    \label{eqn:alignment4color}
\end{equation}

Then, consistent with  the conversion done for PRNU from individual color components to a single combined value in (\ref{eqn:noise_weights}), we compute the weighted average of alignments in each color component and the general form of the RoA, $A$, as:
\vspace{-1mm}
\begin{equation}
\label{eqn:align_weights}
    A=0.3\times A_R + 0.6\times A_G + 0.1\times A_B, 
\end{equation}

\vspace{-1mm}
\begin{figure*}[!ht]
	\centering
    \subfloat[Computation of $I^{Bin}_R(1,2)$ pixel for binning: The left image, $I^{raw}$ is $8\times8$ raw sensor output, the middle image, $I^{temp}$, is obtained after binning, and the right one, $I^{Bin}_R$ is red component of the final $4\times4$ image after demosaicing. The figure shows eight pixels (\textit{i.e.} $(1,1), (1,3),\dots(3,8)$) are used to compute $I^{Bin}_R(1,2)$ pixel index for binning.] {\label{fig:roa_bin2}    
    \includegraphics[trim={1.8cm 20.8cm 5.55cm 1.9cm},clip,width=110mm] {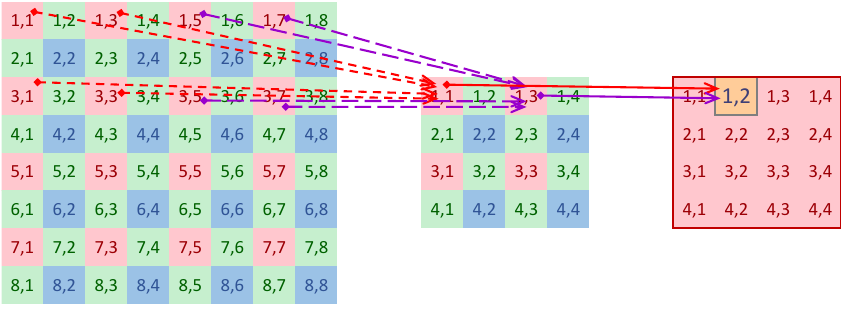}}~
    
    \subfloat[Computation of $I^{Bscale}_R(1,2)$ pixel for bilinear scaling: The left image, $I^{raw}$ is the sensor output, the middle image, $I^{temp}_R$ is obtained after demosaicing applied in camera, and the right one, $I^{Bscale}_R$ is obtained after resizing via bilinear scaling. four pixels are used to compute $I^{Bscale}_R(1,2)$ using bilinear scaling] {\label{fig:roa_bscale}
    \includegraphics[trim={1.8cm 20.8cm 2.5cm 1.9cm},clip,width=128mm] {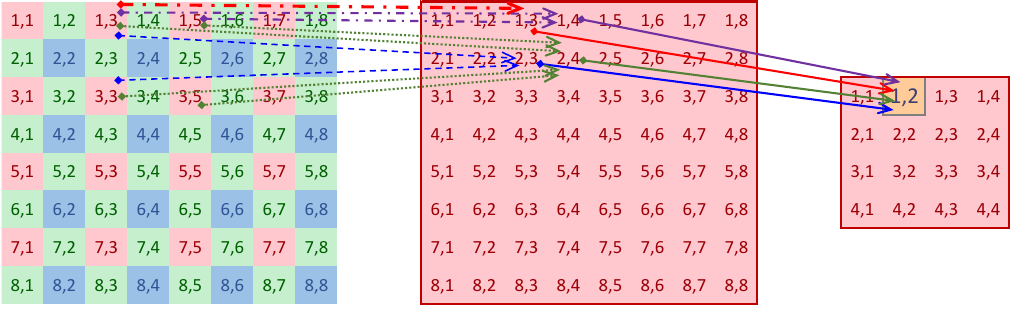}}

\vspace{-1mm}
\caption{Alignment of the same pixel index for binning vs. bilinear scaling}
\label{Fig:align:bin_vs_bilinear}
\end{figure*}

\textbf{Example:} To illustrate, let us take binning (with demosaicing) and bilinear scaling (with demosaicing) as examples of two resizing operations, and find the RoA between the binned media and the bilinearly scaled media shown in Fig.~\ref{Fig:align:bin_vs_bilinear}. Assume that the binned media is a video frame and the bilinearly scaled media is an image. Both these media, however, are captured using an $8 \times 8$ raw camera sensor as shown in  Fig.~\ref{Fig:align:bin_vs_bilinear}. During the in-camera video capturing process, the sensor output is first resized to half using binning, as shown in Fig.~\ref{Fig:binning}, and then the binned output is bilinearly interpolated during demosaicing. The demosaiced output, $I^{Bin}$, is produced as a video frame. It can be seen here that the red color channel of the $(1,2)$ pixel index of $I^{Bin}_R$, $I^{Bin}_R(1,2)$, is computed from eight input pixels of the raw sensor output, $I^{raw}$ (which are shown by the arrows in Fig.~\ref{fig:roa_bin2}). Each input pixel contributes equally to the output pixel. Thus, $I^{Bin}_R(1,2)$ is computed as $I^{Bin}_R(1,2) = \frac{1}{8}(I(1,1)+I(1,3) + ... + I(3,7))$.

In the second scenario, suppose there is another $8 \times 8$ image obtained from the same camera sensor. The image was captured with no in-camera resizing technique and subsequently demosaicing was applied to the sensor output. To match the image (or its PRNU noise) with the video (or its fingerprint), we resize the image (as an out-camera operation) to half resolution using bilinear scaling. To calculate the same pixel index as before (i.e., the red component of $(1,2)$ pixel index of $I^{Bscale}$, $I^{Bscale}_R(1,2)$), we first compute the pixel values which are used to compute it in demosaiced image. As shown in Fig.~\ref{fig:roa_bscale}, four pixels (i.e., $I^{\rm temp}(1,3)$, $I^{\rm temp}(1,4)$,  $I^{\rm temp}(2,3)$, $I^{\rm temp}(2,4)$) in the demosaiced image contributes to $I^{Bscale}_R(1,2)$. The values of these pixels are
\begin{itemize}
\setlength\itemsep{0.15em}
    \item $I^{temp}(1,3) = I(1,3)$
    \item $I^{temp}(1,4) = \frac{1}{2} \big( I(1,3) + I(1,5) \big)$
    \item $I^{temp}(2,3) = \frac{1}{2} \big( I(1,3) + I(3,3) \big)$
    \item $I^{temp}(2,4) = \frac{1}{4} \big( I(1,3) + I(1,5) + I(3,3) + I(3,5) \big)$
\end{itemize}
where $I(m,n)$ are the pixel values of the raw image, $I^{raw}$. Therefore, by averaging these four pixels of $I^{temp}$, we can compute the weights of all the input pixels creating $I^{Bscale}_R(1,2)$ as $I^{Bscale}_R(1,2) = \frac{9}{16} I(1,3)+ \frac{3}{16}\big[I(1,5) + I(3,3)\big]+\frac{1}{16}(I(3,5))$.

Using (\ref{eqn:alignment4px_new}), the alignment of the red component at pixel index $(1,2)$ between $I^{Bin}_R(1,2)$ and $I^{Bscale}_R(1,2)$ \Big(i.e., $a_R\big(I^{Bin}(1,2), I^{Bscale}(1,2)\big)$\Big) can be found as {\color{black} $a_R\big(I^{Bin}(1,2), I^{Bscale}(1,2)\big) = \frac{1}{8} + \frac{1}{8} + \frac{1}{8}+ \frac{1}{16} = \frac{7}{16}$}. Similar analysis can be done for the other pixel positions in all three color planes and a more precise characterization of the alignment between the two scaled images can be made. 

Having established notation and the definition of RoA with the aid of examples, in the subsequent sections we derive RoA for a bilinearly scaled image by half, with images scaled to the same size but using binning and line-skipping respectively. %To do this, in the next sub-section, we first derive the sensor-pixel correspondences of  the bilinearly scaled image. That is the set of sensors corresponding to the original image that contribute to each pixel in the bilinearly scaled image. Then, in the next two-subsections, we derive similar correspondences for binned, and line-skipped images and based on the earlier derived correspondences for bilinearly scaled images, compute the RoA for each case.
For brevity of analysis, we assume that in each case scaling is done from a full resolution image which represents the actual resolution of the underlying sensor. We also assume that scaling is done by exactly half the dimension of the full resolution sensor. The analysis presented is valid for any Bayer pattern (i.e., $RGGB$, $BGGR$, $GRBG$, and $GBRG$) as each of them results in the same ratio of alignment. %Also, analysis for red color is similar to the analysis for blue color. Therefore, it is done for red and green colors only. 
Also for brevity, the analysis is provided only for red color. Analysis of green and blue colors are similar to red. %{\color{black}For green color, the analysis is provided in Appendix~\ref{apdx:line}}. 

Besides, many cameras do binning before linearization; that is, they average the electrical charges in pixels instead of intensity which helps decrease read-out and shot noise. However, these noise sources have limited effect on this analysis; hence we simply consider binning is done by taking the average of intensity values of each pixel being accumulated.

Finally, it is crucial to note that one of the main differences between binning and line-skipping (\textit{i.e., in-camera resizing}) and bilinear scaling (\textit{i.e., out-camera resizing}) is that binning and line-skipping are done before demosaicing (color filter array interpolation) whereas the bilinear scaling is done after demosaicing. Both binning and bilinear scaling use bilinear interpolation, however, the way they use it differ as shown in Fig.~\ref{Fig:align:bin_vs_bilinear}. For binning, interpolation is applied on the Bayer filter output for which a composite pixel is produced from four same-color neighboring pixels. Interpolation in bilinear scaling is applied on the final RGB image from the sensor after downsizing each color channel separately.

\vspace{-2mm}
\subsection{Sensor-pixel correspondence for bilinear scaling} 
\vspace{-1mm}
\label{sec:math:bilinear}
We consider the case when the higher resolution media is downsized using bilinear scaling after the media has been captured (i.e., out-camera resizing). This is done when an image fingerprint is being matched to a video fingerprint or vice versa.  {\color{black} We assume bilinear scaling is implemented using  $\frac{1}{4}\bigl[ \begin{smallmatrix}1 & 1\\ 1 & 1 \end{smallmatrix}\bigr]$ downsampling kernel. In the demosaicking step, for red and blue components the convolution kernel, $h_{r,b} = \frac{1}{4}\Bigl[ \begin{smallmatrix}1 & 2 & 1\\ 2 & 4 & 2 \\1 & 2 & 1 \end{smallmatrix}\Bigr]$, and for green channel $h_{g} = \frac{1}{4}\Bigl[ \begin{smallmatrix}0 & 1 & 0 \\ 1 & 4 & 1 \\0 & 1 & 0 \end{smallmatrix}\Bigr]$ are used which do bilinear interpolation. Although these are basic filters and more complicated kernels may be used in real cameras, for the sake of simplicity of analysis, we used these kernels.}

As we have discussed before, bilinear scaling is done on the demosaiced sensor output. When the raw image, $I^{raw}_R$ is demosaiced without any down-sampling operation, the red color component of the output image, $I^{temp}_R(m, n)$, becomes
\vspace{-1mm}
\begin{equation}
\begin{aligned}
    &I^{temp}_R(m, n) = 
    \begin{cases} 
    {\scriptstyle I(2m-1,2n-1)}                 &\text{case 1} \\
        \frac{I(2m-1,2n-1:2:2n+1)}{2}          &\text{case 2} \\
        \frac{I(2m-1:2:2m+1,2n-1)}{2}          &\text{case 3} \\
        \frac{I(2m-1:2:2m+1,2n-1:2:2n+1)}{4}   &\text{case 4}   
    \end{cases} 
    \label{eqn:cfa:red}
\end{aligned}
\end{equation}

Now, suppose bilinear scaling is applied on the demosaiced sensor output to resize it to half resolution. Then, $I_R^{Bscale}(m,n)$, denoting the red color component of the $(m, n)^{\rm th}$ pixel of the scaled output image, can be found as
\vspace{-1mm}
\begin{equation}
\small
    \begin{aligned}
    I_R^{Bscale}(m,n) = \frac{1}{4} \Big( & I^{temp}_R(2m-1,2n-1) + I^{temp}_R(2m-1,2n) + \\ 
                   & I^{temp}_R(2m,2n-1) + I^{temp}_R(2m,2n) \Big)
    \end{aligned}
    \label{eqn:bilinear:red}
\end{equation}
By combining (\ref{eqn:cfa:red}) and~(\ref{eqn:bilinear:red}), we obtain $I_R^{Bscale}(m,n)$ as
\vspace*{-1mm}
\begin{equation}
\small
\begin{aligned}[b]
     I_R^{Bscale}(m,n) = &\frac{9}{16} I(2m-1,2n-1) + \frac{1}{16} I(2m+1,2n+1) + \\
                         &\frac{3}{16} \Big(I(2m-1,2n+1) + I(2m+1,2n-1) \Big).
    \end{aligned}
\label{eqn:bilinear:red:final}
\end{equation}

Using a same approach, $(m,n)^{\rm th}$ pixel of $I_G^{Bscale}$ can be obtained {\color{black}(details can be found in Appendix~\ref{apdx:bilinear})}:
\vspace*{-1mm}
\begin{equation}
\small
    \begin{aligned}
        I_G^{Bscale}(m,n) = &\frac{3}{8} \Big(I(2m-1,2n)+I(2m,2n-1)\Big) + \\ 
                            &\frac{1}{16} \Big(I(2m-2,2n-1)  + I(2m-1,2n-2) + \\
                            & ~~~~~~I(2m,2n+1)+I(2m+1,2n)\Big).
    \end{aligned}
    \label{eqn:bilinear:green:final1}
\end{equation}

So,~(\ref{eqn:bilinear:red:final}) and~(\ref{eqn:bilinear:green:final1}) show sensor-pixel correspondence for the red and green color planes when the raw image, $I^{raw}$ is resized by half via bilinear scaling, $I^{Bscale}$. In the next subsections, we will obtain the sensor-pixel correspondence for images resized to half size by binning, $I^{Bin}$, and line-skipping, $I^{Line}$, and then find their RoA with $I^{Bscale}$.

%So,~(\ref{eqn:bilinear:red:final}) shows the sensor-pixel correspondence for the red  color planes when the raw image, $I^{raw}$ is resized by half via bilinear scaling, $I^{Bscale}$. In the next subsections, we will obtain the sensor-pixel correspondence for images resized to half size by binning, $I^{Bin}$, and line-skipping, $I^{Line}$, and then find their RoA with $I^{Bscale}$

\vspace{-5mm}
\subsection{Sensor-pixel correspondences for binning}
\label{sec:math:bin}
Similar to the above section (i.e., Section~\ref{sec:math:bilinear}), we can obtain the red and green components of a still image resized with the binning approach we described in Section~\ref{sec:background}. Considering the differences of binning and bilinear scaling, we can obtain the $(m,n)^{\rm th}$ pixel of the red  component of an image resized with binning, $I_R^{Bin}(m,n)$:

\vspace*{-1mm}
\begin{equation}
    I_R^{Bin}(m,n) = 
    \begin{cases}
        \frac{I(2m-1:2:2m+1,2n-1:2:2m+1)}{4}       &\text{case 1} \\
        \frac{I(2m-1:2:2m+1,2n-3:2:2n+3)}{8}       &\text{case 2} \\
        \frac{I(2m-3:2:2m+3,2n-1:2:2n+1)}{8}       &\text{case 3} \\
        \frac{I(2m-3:2:2m+3,2n-3:2:2n+3)}{16}      &\text{case 4}.\\
    \end{cases}
    \label{eqn:binning:red:final1}
\end{equation}
and green component $I_R^{Bin}(m,n)$ as:
\vspace*{-1mm}
\begin{equation}
    \begin{aligned}
    &I_G^{Bin}(m,n) = \\ 
    &\begin{cases}
        \frac{I(2m\tmin4:2:2m+2,2n\tmin1:2:2n+1)+I(2m\tmin1:2:2m+1,2n\tmin4:2:2n+2)}{16} &\text{case 1} \\
        \frac{I(2m\tmin1:2:2m+1,2n\tmin2:2:2n)}{4}                              &\text{case 2} \\
        \frac{I(2m\tmin2:2:2m,2n\tmin1:2:2n+1)}{4}                              &\text{case 3}\\
        \frac{I(2m\tmin3:2:2m+3,2n\tmin2:2:2n)+I(2m\tmin2:2:2m,2n\tmin3:2:2n+3)}{16}    &\text{case 4}.\\   
    \end{cases}
    \end{aligned}
    \label{eqn:binning:green:final}
\end{equation}
{\color{black}details of which can be found in Appendix~\ref{apdx:binning}.}

\vspace{-5mm}
\subsection{Sensor-pixel correspondences for line-skipping}
\label{sec:math:line}
Line-skipping can be implemented in numerous ways {\color{black}and different implementations may use completely different sets of pixels}. For our anlaysis, we consider that line-skipping is implemented by removing every $2 \times l + 3$ and $2 \times l + 4$ (where $l$ is a natural number) rows and columns from the sensor output (as shown in Fig.~\ref{Fig:lineskip}). In other words, every $3^{\rm rd}$ and $4^{\rm th}$ rows and columns are skipped. Using the same approach as above, we can obtain pixel values of the red component of a still image resized by line-skipping, $I_R^{Line}(m,n)$, as
\begin{equation}
\small
    I_R^{Line}(m,n) = 
    \begin{dcases}
        I(2m-1,2n-1)                             &\text{case 1}  \\ 
        \frac{I(2m-1,2n-3:4:2n+1)}{2}            &\text{case 2} \\ 
        \frac{I(2m-3:4:2m+1,2n-1)}{2}            &\text{case 3}\\ 
        \frac{I(2m-3:4:2m+1,2n-3:2:2n+1)}{4}     &\text{case 4}.\\    
    \end{dcases}
    \label{eqn:line:red}
\end{equation}

Similarly, the green component of the same, $I_G^{Line}(m,n)$, can be found as
\begin{equation}
\begin{aligned}
    & I_G^{Line}(m,n) = \\ 
    &\begin{cases}
        \frac{I(2m-4:4:2m,2n-2)+I(2m-2,2n-4:4:2n)}{4}      &\text{case 1} \\ 
        {\scriptstyle I(2m-1,2n)     }                      &\text{case 2} \\
        {\scriptstyle I(2m,2n-1)       }                   &\text{case 3} \\
        \frac{I(2m-3:4:2m+1,2n-2)+I(2m-2,2n-3:4:2n+1)}{4}  &\text{case 4}.\\   
    \end{cases}
    \label{eqn:line:green}
\end{aligned}
\end{equation}
{\color{black}(For more details, see Appendix~\ref{apdx:line})}

\vspace{-3mm}
\subsection{{\color{black}RoA's of different combinations}}
In the above sections, we show how pixel sensor correspondence can be computed for different resizing schemes. In this section, we compute the RoA between the different resizing schemes.  As stated above, each pixel in the resized image can be computed using one of four different cases based on its index (i.e., whether odd or even). We assume the occurrence of each case is equal (the difference in occurrences are negligible for high-resolution images). Therefore, averaging the alignment of these four cases will yield the RoA of the whole image. Also, as clarified before, the RoA computations of red and blue color planes will be the same. For the green color plane, the computation will differ.

Suppose $A_R(I^{Bin}, I^{Bscale})$ denotes the RoA of the red color component (for all pixels) between a binned image and a bilinear scaled image. Then, $A_R(I^{Bin}, I^{Bscale})$ can be found by averaging the alignments obtained for the $4$ cases mentioned in Section~\ref{sec:Syntax} (i.e., case $1$, \dots case $4$).

We can obtain the RoA for red color of the two output images (i.e., $I_R^{Bin}$ and $I_R^{Bscale}$) by~(\ref{eqn:bilinear:red:final}), and~(\ref{eqn:binning:red:final1}). For example, red component of case 1 in binning is
$$\frac{I(2m-1:2:2m+1,2n-1:2:2m+1)}{4} $$ 
whereas any pixel in bilinear scaling is

\begin{equation*}
\small
\begin{aligned}
&\frac{9}{16} I(2m-1,2n-1) + \frac{1}{16} I(2m+1,2n+1) + \\
&\frac{3}{16} \Big(I(2m+1,2n-1) + I(2m-1,2n+1)\Big)
\end{aligned}
\end{equation*}

So, using \ref{eqn:alignment4px_new}, their alignment for case 1, $A_R^{Bin}(\text{case 1})$ becomes $min(\frac{1}{4}, \frac{9}{16}) + min(\frac{1}{4}, \frac{3}{16}) + min(\frac{1}{4}, \frac{3}{16}) + min(\frac{1}{4}, \frac{1}{16}) = \frac{11}{16}$ (i.e., the minimum weights of the common pixels $I(2m-1,2n-1)$, $I(2m-1,2n+1)$, $I(2m+1,2n-1)$, and $I(2m+1,2n+1)$, respectively).

When we calculate the alignment of red component using \ref{eqn:alignment4px_new}, we can obtain $A_R(I^{Bin}, I^{Bscale})$ as
\begin{equation}
\small
\label{eqn:red_percentage}
   A_R(I^{Bin}, I^{Bscale}) = \frac{1}{4} \times (\frac{11}{16}+ \frac{7}{16}+ \frac{7}{16}+ \frac{4}{16})= \frac{29}{64}.
\end{equation}

Similarly, we can obtains the alignment for green channel using ~(\ref{eqn:bilinear:green:final}) and~(\ref{eqn:binning:green:final}). 

\begin{equation}
\small
\label{eqn:green_percentage}
    A_G(I^{Bin}, I^{Bscale}) = \frac{1}{4} \times (\frac{6}{16}+ \frac{6}{16}+ \frac{6}{16}+ \frac{12}{16}) = \frac{30}{64}.
\end{equation}

Note that alignment of blue color  $A_B^(I^{Bin}, I^{Bscale})$ is also $\frac{29}{64}$. Thus by using~(\ref{eqn:align_weights}), the RoA of the whole image can be computed as 
\begin{equation*}
 \begin{aligned}
 A(I^{Bin}, &I^{Bscale}) = 0.3\times A_R(I^{Bin}, I^{Bscale}) + \\  
 &0.6\times A_G(I^{Bin}, I^{Bscale}) + 0.1\times A_B(I^{Bin}, I^{Bscale}).
 \end{aligned}  
\end{equation*}
Putting values of alignment for each color component in this equation, we can get $A(I^{Bin}, I^{Bscale})=0.4625$. 

The RoA of other cases (such as $I^{Line}$ and $I^{Bin}$ etc) can be found using the same approach. {\color{black}Table~\ref{tab:roa:all} shows the possible combinations of  these $3$ resizing approaches. If two images are resized with the same resizing technique, their RoA will be $1.00$. But when they are resized using different techniques, their RoA will decrease based on the extent of the common sensor elements contributing to each pair of spatially corresponding pixels in the resized images.}

\vspace{-1mm}

\begin{table}[h]
\centering
\caption { RoA for media resized differently}
{\color{black}\begin{tabular}{|c|c|c|c|}
\hline
Train\textbackslash{}Test & Bscale  & Bin & Lskip   \\ \hline
Bscale                    & 1.00   & 0.46   & 0.17    \\ \hline
Bin                       & 0.46   & 1.00   & 0.21    \\ \hline
Lskip                     & 0.17   & 0.21   & 1.00    \\ \hline
\end{tabular}}
\label{tab:roa:all}
\end{table}

{\color{black}From the RoA calculation, we can infer that when a video is resized via binning and the image FE via bilinear scaling with the same factor, the video FE and image FE  can still match since there is significant alignment between the two. Line skipping, however results in lower RoA in the case of a mismatch. This could be due to the fact that line skipping entirely discards pixels whereas binning and bilinear scaling compute a composite pixel, parts of which may still align between the resized image FE and video FE. However, for some edge cases (e.g., when the correlation value is slightly below the decision threshold), matching by resizing with another (the correct) scheme might yield a match decision. Hence using a single resizing technique, may not be the best option. This insight is used to develop a matching algorithm in the next section.}

\vspace{-3mm}
\subsection{Experimental validation}
To evaluate the impact of RoA on correlation, we simulated an experiment which calculates $\rho$ (Pearson Correlation Coefficient) and True Positive Rate (TPR) when different resizing techniques are applied on train and test images. 
We used the \textit{Raise} dataset which contains a set of RAW images provided by Dang-Nguyen et al. \cite{dang2015raise}.
From this dataset, we obtained $500$ images: $100$ for training and $400$ for testing. We resized each image to half size via (i) bilinear scaling, (ii) binning, and (iii) line-skipping {\color{black} We implemented these three methods as described in Section~\ref{sec:math:bilinear} and~\ref{sec:math:bin}). After resizing the images with each  of these methods we obtain three copies of each image. From the training images, we extract three camera fingerprints and three PRNU noise patterns from the test images. 
We then correlate each fingerprint with each PRNU noise. This way, we do nine different correlations.

Table~\ref{tbl:math:sim:corr} shows that the average correlation, $\overline{\rho}$, for the different resizing cases. In this table, rows and columns indicate how the training and test images are resized, respectively. 

\vspace{-1mm}

\begin{table}[h]
\centering
\caption {Average correlation for media resized differently}
{\color{black}\begin{tabular}{|c|c|c|c|}
\hline
Train\textbackslash{}Test & Bscale & Bin & Lskip    \\ \hline
Bscale          & 0.0308 & 0.0060 & 0.0033          \\ \hline
Bin             & 0.0064 & 0.0186 & 0.0127          \\ \hline
Lskip           & 0.0038 & 0.0130 & 0.0344          \\ \hline
\end{tabular}}
\label{tbl:math:sim:corr}
\end{table}

Similarly, Table~\ref{tbl:math:sim:tpr} shows the TPR of these combinations (i.e., TPR is obtained using PCE with a threshold of $60$). As can be seen,  RoA is aligned with both $\overline{\rho}$ and TPR. When the RoA is $1.00$ (i.e., both training and test images are resized with the same technique), $\overline{\rho}$ is high which results in TPRs that are above $90\%$ in all cases. When RoA decreases, correlation also decreases which leads to a lower TPR. Interestingly, when  either training or test images are resized with binning and the other with line-skipping, it achieves better TPR compared to bilinear scaling vs line-skipping.}

\begin{table}[h]
\centering
\caption {{\color{black}TPR for media resized differently}}
{\color{black}\begin{tabular}{|c|c|c|c|}
\hline
Train\textbackslash{}Test & Bscale & Bin & Lskip  \\ \hline
Bscale                    & 0.95 & 0.66 & 0.50    \\ \hline
Bin                       & 0.69 & 0.90 & 0.82    \\ \hline
Lskip                     & 0.55 & 0.84 & 0.98    \\ \hline
\end{tabular}}
\label{tbl:math:sim:tpr}
\end{table}

Note that although the correlation significantly depends on the RoA, the contribution of individual pixels to the PRNU noise is also another factor. This contribution depends on image content and quality, and hence is difficult to model. Further, the analysis is for an idealized case where the down-sampled video is half the original sensor resolution. But the analysis is valid as it gives insight into the relative performance achieved for attribution in the presence of mixed media when different in-camera capture techniques are used. {\color{black} Also, it should be noted that we did the analysis for bilinear scaling, however, the same calculation can be done for bicubic or Lanczos scaling. When we calculate the RoA of bilinear and bicubic scaling (or other scaling methods), they are typically very high (i.e., $>0.9$) as they do similar processing.}

Finally, it should also be noted that there are many operations that are performed within the camera such as JPEG compression, denoising, or gamma correction. Each of them can also play a role in PRNU attribution performance. However, the focus of this work is PRNU attribution in the presence of misalignment. Since binning, line-skipping, and demosaicing are the only in-camera operations that could potentially cause misalignment between sensors and pixels, and we use bilinear scaling, we considered only these operations in the mathematical analysis presented. The rest of the in-camera processing were not included as they do not directly contribute to misalignments. 

\vspace{-4mm}
\section{Camera Attribution with Mixed Media}
\label{sec:approach}
Now that we have a better understanding of the different ways resizing may be done within a camera and the RoA between them, we present a generic algorithm for source camera identification between images and videos. 

%In this section, we look at different scenarios under which camera attribution may need to be done with different types of visual objects. Specifically, we look at the following cases:
 
%\begin{itemize}
 %\item Matching the fingerprint computed from a set of images from a known camera with a query video. %This general scenario has multiple possibilities. Although the image fingerprint is available, the camera model may not be known as the images may have been obtained from a social media platform such as Facebook. Along with that, 
 %the resolution and/or the aspect ratios of the visual media objects may be different. %Similar to images, the video header may not contain the camera model information. 
 
 %\item Matching a video fingerprint with one or more images of unknown origin. This case is the opposite of the above case. %Similar to the above case, the video and image(s) may not have the associated metadata. 
% \item Although our main focus is mixed media, for the sake of completeness, we include the case when both query and test media are videos whose resolutions and aspect ratios might be different as the attribution of such media has not been addressed yet in the literature. 
%\end{itemize}

%Below, we provide solutions for each of the above scenarios.

%\vspace{-3mm}
%\subsection{Mixed media}
% \label{sec:our_approach:mixed}
The solution for source camera attribution is independent of whether it is the images or the videos that are from the known source camera. Hence, in the rest of this section, diverting from convention, and to avoid confusion among the different scenarios, we refer to all noise patterns and fingerprints obtained from images and videos as \textit{image FE (fingerprint estimate)} and \textit{video FE} respectively. We do this even when the estimate is obtained using just a single image or from a very short video. 

{\color{black}
To focus better on our contribution, we assume that neither the video FE nor the image FE is obtained from media that are zoomed, stabilized, or obtained using a non-linear operation (such as HDR). Moreover, we assume they were not subjected to out-camera cropping and/or resizing operation. There has been considerable research lately to perform attribution in the presence of such operations. For example, recent research has led to techniques to obtain a camera fingerprint from stabilized video \cite{taspinar2016source, iuliani2017hybrid, mandelli2018facing}. Similarly, Goljan \cite{Miroslav:Private} has proposed a way to obtain camera fingerprints from HDR images. Also \cite{gallagher2005detection} and \cite{mahdian2008blind} use JPEG compression artifacts or periodic interpolation artifacts to find zooming factor from images. So, given that these operations are somewhat ``accurately'' reversed in visual media, they can be applied to obtain the fingerprint estimates before the proposed algorithm is used. Of course, attribution performance may drop in some cases, and parameters of the proposed algorithm (such as search range) will have to be re-adjusted in other cases.

Noted that our assumptions do not restrict the applicability of the techniques presented. As will be shown in the next section, experiments show that the proposed algorithm gave good performance even when the test set included data that were subjected to zooming, cropping, and stabilization.}

\tikzstyle{block} = [draw, rectangle, minimum height=1.6cm, minimum width=2.4cm]
\tikzstyle{blockbound} = [draw, rectangle, minimum height=4em, minimum width=6em, node distance=4cm, line width = 1.5mm]
\tikzstyle{sum} = [draw, fill=blue!20, circle, node distance=1cm]
\tikzstyle{relationship} = [top color=white, bottom color=red!10,
                                  draw=red!50!black!100]

\tikzstyle{paralleloid}=[trapezium,trapezium stretches=true, draw, minimum height=1cm, minimum width=3cm, top color=white, bottom color=blue!15, 
trapezium left angle=60, trapezium right angle=120]

\definecolor{forestgreen}{RGB}{34,139,34}

\begin{figure}[]
\scalebox{0.75}{
\begin{tikzpicture}[node distance=2cm,>=latex', every text node part/.style={align=center}]
    \node [block, minimum height=2.25cm, minimum width=3cm] (imFE) {image FE};
    \node [blockbound, minimum height=1.75cm, minimum width=2.5cm, below=1cm of imFE, xshift=-0.2cm] (vidFE) {video FE};
    
    \node [paralleloid, right=1.35cm of imFE, yshift=-0.6cm] (find) {find the\\search range}; 
    \node [block, right of=imFE, node distance=8cm] (resize) {resized\\image FE}; 
    \node [block, right= 1.9cm of vidFE] (center) {center of\\video FE}; 
    \node [relationship, diamond, right=0.4cm of center, aspect=1.4] (ncc) {NCC(video FE,\\resized image FE)};

    \node [block, fill=forestgreen, text=white, below =1cm of ncc] (match) {MATCH};

    \node [relationship, diamond, below=1.3cm of center, aspect=1.4] (more) {more resizing\\ techniques?};

    \node [block, fill=red!150, text=white, left= 1.4cm of more] (mismatch) {MISMATCH};

% --------------------------------------ARROWS -----------------------------
    \draw [->] (vidFE) -- node[anchor=north, text width=2cm] {crop\\boundary\\pixels} (center);
    \draw [->] (imFE) -| ([shift={(7mm,3mm)}]imFE.east) -| ([shift={(-7mm,3mm)}]resize.west) -| ([shift={(-7mm,3mm)}]resize.west) -| ([shift={(-7mm,0mm)}]resize.west) -- node[anchor=south,xshift=-2.2cm, yshift=0.25cm] {choose a resizing technique} (resize);
    \draw [->] (center) -- node  {}(ncc);
    \draw [->] (resize) -- node  {}(ncc);
    
    \draw [->] (center) -- node  {}(find);
    \draw [forestgreen,->] (ncc) -- node [anchor=west, yshift=-1mm] {pce $> \tau$}(match);
    
    \draw [->,red] (more) -- node [anchor=north] {No}(mismatch);

    \draw [->,PineGreen, very thick] (more) |- ([shift={(6.2cm,-2mm)}]more.south) node [anchor=north, xshift=-3.2cm]  {Yes: reset the search and\\choose next technique} -| ([shift={(1.2cm,0.7cm)}]resize.east)  |- ([shift={(0cm,1.35cm)}]find.north) -| ([shift={(0cm,1cm)}]find.north);

    \draw [->] (find) -| ([shift={(6.7mm,6mm)}]find.east) |- node  {}(resize);
    
    \draw [->] (imFE) -| ([shift={(7mm,-6mm)}]imFE.east) |- node  {}(find);
    \draw [->,red] (ncc) -| ([shift={(1mm, 11mm)}]ncc.east) |- node [anchor=east,yshift=-1.5cm]{ pce $<\tau$, \\try next \\factor }(resize);

    \draw [->,red] (ncc) |- ([shift={(-12mm, -3mm)}]ncc.south) -| node [anchor=north, xshift=2cm]{ end of search }(more);
    
    \draw [blue, very thick, dashed](2.9,-1.8) -- (2.9,-4.7) -- (10.3,-4.7) -- (10.3, 1.1) -- (6.6, 1.1) -- (6.6, -1.8) -- cycle;
    
\end{tikzpicture}
}

\caption{Correlation of an image FE with a video FE}
\label{fig:exh_search}
\end{figure}
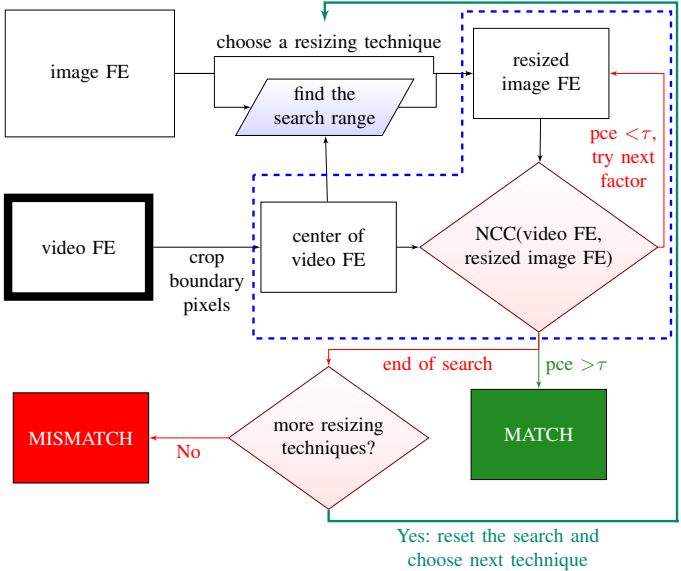

Fig.~\ref{fig:exh_search} gives a summary of the proposed algorithm for source camera attribution, that leverages the knowledge and insights obtained from the different in-camera resizing operations presented in Section 2 and RoA results in Section 3. 

\vspace{-2mm}
{\color{black} \paragraph{Step 1:} The fingerprint computed from the video is cropped by removing $2\times h$ columns along the height ($h$ columns each from the left and the right borders) and $2 \times w$ rows from the width. This step is required to overcome any misalignment that can potentially arise due to the use of boundary pixels for capturing a video which may not have been used while capturing an image as described in Section 2. Specifically, it can result in the computation of a different search range in Step 4 as will be explained later. The values of $h$ and $w$ vary with camera models. They can, however, set to a maximum number for making sure that the boundary pixels over all possible camera models have been cropped out. 

\vspace{-2mm}
\paragraph{Step 2:} Now we select a candidate resizing technique to use. As described in Section~\ref{sec:background}, binning, line-skipping and bilinear scaling may use different sets of pixels in the raw image to obtain the same resolution video frame. As analyzed in the Section~\ref{sec:math}, the use of different resizing techniques causes a decrease in RoA which may result in failure to match in some cases even if the two media are resized with the same factor. 
Therefore, in contrast to the belief that scaling would be sufficient to match a video FE with an image FE, different resizing techniques may have to be tried. This could be very time consuming as the techniques themselves have different parameters. Scaling could use different kernels. Binning can be done in different ways depending on the Bayer pattern being used, and line-skipping has numerous implementation possibilities. Based on the results in Section~\ref{sec:math} (both analytical and experimental) we see that bilinear scaling and binning appear to provide better RoAs as opposed to line-skipping. Hence in the interest of efficiency, we propose the following resizing technique selection strategy. When we have good estimates of both the camera and the video fingerprint, just bilinear scaling can be employed. Otherwise, first, try bilinear scaling. If there is no match, then try four different combinations of $2 \times 2$ binning. 

%we believe that either scaling or binning may be required when matching a video FE with an image FE (or vice versa). With this observation, we first resize the query image using bilinear scaling. If there is a mismatch (i.e., the image FE did not match with the video FE), we then apply the following steps on the query image: we first revert RGB image to raw, then downsize to half with $2\times2$ binning scheme. We then consider the binned image is a raw image and apply basic imaging pipeline on it. 
%For binning, we consider all four Bayer patterns for best possible matching. 

%Notice that for images which were already downsized within a camera, this scheme is expected not to work as there is an irreversible interpolation on them.

\vspace*{-1mm}
\paragraph{Step 3:} In this step we determine the search range of the possible resizing factors that need to be tried to perform the match. To accurately determine the search range, we have to take multiple factors into account including the video and image resolutions, possible boundary pixels issue (which we crop in Step 1), the in-camera resizing techniques and difference in aspect ratios, if any. In the next subsection, we describe how to compute the correct search range. 

\vspace*{-1mm}
\paragraph{Step 4:} Although the search range specifies the different resizing factors to try, not all are equally likely. So instead of just starting from the lowest and working ones way to the highest (as is current practice), based on the knowledge gained from Section 2 and our experiments, we propose an ordering of resizing factors to be used by first trying the more likely factors. Details are provided in the next subsection. 
 
\vspace*{-1mm}
\paragraph{Step 5:} The scaled image FE is correlated with the cropped video FE using NCC. If the PCE result at the NCC peak is above a threshold, it is concluded that the image(s) and video are taken by the same camera and the algorithm halts. Otherwise, the next resizing factor is tried. When all the different resizing factors in the search range have been tried, and no match is found, then we go back to step 2 to try a different resizing technique and repeat steps 4 and 5. If no more resizing techniques are available, then the media objects are considered to have been taken by a different camera.
}

\vspace*{-5mm}
% ------------------------------------------ SMART SEARCH ------------------------------
{\color{black}
\subsection{Smart search} 
\label{SmartSearch1} 
% --- this is step 3 (finding search range) --- and step 4(finding the minimum possible search range)

% step 3
%In this subsection, we explain \textit{smart search} which corresponds to the Step 3 and 4 in the above algorithm. Here, it is crucial to determine the correct range of resizing factors that the video might have been resized with. Failing to determine it might results in a low performance.

% this part is step 4
%On the other hand, the algorithm above needs to explore different resizing techniques and resizing factors, it is essential to find ways to choose minimum possible search range such that we can speed up the search. We describe some simple heuristics that provide up to 5 times speedup over a naive exhaustive search. This involves first narrowing down the range of resizing factors that are searched and the second involves trying them in decreasing order of likelihood. That is, trying the most likely factors within the range first. 

 Since the algorithm above needs to explore different resizing techniques and resizing factors, it is important to find ways to speed up the search. In this subsection, we describe some simple heuristics that provide up to 5 times speedup over a naive exhaustive search. This involves first narrowing down the range of the resizing factors that are searched and the second involves trying them in decreasing order of likelihood. That is, trying the most likely factors within the range first. 

\paragraph{Determining the search range} 
Suppose the resolution of the video is $m\times n$ and of the image is $p \times q$. Suppose also that the image is resized with a factor, $r_1$ within the camera and the video with $r_2$. Our goal is to find the values of $r_1$ and $r_2$ such that we can match the video FE to the image FE.%, but when $r_1\neq 1$, it may ...

Let us first assume that we know $r_1$. If this is the case, we can determine the in-camera resizing factor of the video, $r_2$, by considering the following two cases:

\begin{enumerate}[label=(\roman*),font=\itshape]
\item The aspect ratio of the image is the same as that of the active image region of the sensor.
\item The aspect ratio of the image is different from the active image region. 
\end{enumerate}

\textit{Case (i):} %In this case, the image might have been resized with a factor, $r_1$, within the camera; however, its aspect ratio remained the same as full sensor resolution. 
For this case, the search range for $r_2$ becomes $\bigl[max(m/p, n/q), 1/{r_1}\bigr]$. Notice that this search range is similar to the one proposed by Goljan et al. \cite{jessica:crop} with the addition of the $r_1$ variable in the equation. To clarify with an example, suppose the image is captured at a resolution of $4000\times3000$ using the active image region of the sensor and the Full HD video ($1920\times1080$) is captured from the same region without using the active boundary. Notice that the resolution (and aspect ratio) of the image is the same as the active image region and hence the resizing factor, $r_1$ is $1$. In this circumstance, the search range becomes $[max(1920/4000, 1080/3000), 1/1] = [0.48, 1]$. So, we iteratively resize the image FE with these resizing factors and correlate with the video FE. Now suppose that the image was resized with $r_1=0.5$. So its resolution became $2000\times1500$. In this case, the search range becomes $[max(1920/2000, 1080/1500), 1/0.5] = [0.96, 2]$. Without $r_1$, only the search range of [$0.96, 1$] would be explored. However, this range is likely to fail. %For example, when the video is captured with active image and then resized to a resolution greater than $2000\times1125$ and cropped, the algorithm in \cite{jessica:crop} will fail.

% \textbf{ADD JESSICA'S HERE ALGORITHM HERE}
% sensor/image:16x12 
% video: 12x9 (video is resized 3/4) 

% now lets say image is 12x12 search range [1, max(m/p, n/q)] = [1,1]
% normally it should have been [1,3/4] with [1, min(m/p, n/q)] = [1,3/4]

\textit{Case (ii):} This case is not as intuitive as case (i). Here, when the aspect ratio of the image is different from the sensor's, choosing $max(m/p, n/q)$ as the lowest resizing factor may become inaccurate. To understand this better, consider the following example. Suppose the image resolution is $1600\times1200$ as captured from only the active image and no resizing ($r_1=1$). The video is also captured from the active image region and then downsized to $1200\times900$ (i.e., $r_2=3/4$). Using the formula in case (i), the search range becomes $[1,3/4]$ which includes the correct resizing factor of the video. Now suppose, another image whose resolution is $1200\times 1200$ is taken by the same camera with only a part of the active image region without any scaling operation, only cropping from the sides (i.e., $r_1=1$). If we use the formula in case (i), the search range will be $[1,1]$ which does not contain $3/4$, so the search will fail. To fix this, we need to change the lower bound of the above formula to $\bigl[min(m/p, n/q), 1/{r_1}\bigr]$. This results in the range $[1,3/4]$ and the search ends in a correct match. It is crucial to see this is one of the main differences separating the proposed approach from \cite{jessica:crop} in terms of finding the search range. 

Note that the search range calculation is done disregarding the boundary pixel issue. Recall that in step 1 we crop $20$ rows from top and bottom of the video FE and appropriate number of columns that maintains its aspect ratio. For example for a $4:3$ video, $15$ columns from right and left of the video. This way, we handle the boundary pixel issue without changing the search range. 

% This is because changing aspect ratio will change the ratio of $p$ or $q$, therefore maximum of $m/p$ and $n/p$ may be higher than the actual resizing factor of the video, $r_2$.Therefore, we update the above search range as follows, $\bigl[min(m/p, n/q), 1/{r_1}\bigr]$. 

We have examined the case where $r_1$ is known. This, for example, is true when the image FE is obtained with images captured at full resolution, giving $r_1 = 1$, as is often the case. By full resolution, we mean the resolution of the active image region of the sensor as shown in Fig.~\ref{Fig:boundary}. However, when the image FE is obtained from images captured with a lower resolution due to in-camera resizing (as in Fig.~\ref{Fig:l2}), we can find $r_1$ by searching across low and full resolution images dimensions captured from a camera of the same model. If a full resolution image from the same camera is not available, choosing the minimum possible resizing factor (i.e., number of rows and columns of low-resolution image divided by the rows and columns of the full resolution image, respectively) will work.

{\color{black} Finally, it should be noted that search by also be sped up using the approach proposed by \cite{jessica:crop}, where it has been shown that even if a query image is resized by a factor $x$ and the camera fingerprint is by $x \mypm \epsilon$, there still can be effective correlation between them. This is because the RoA between the query image FE and the fingerprint is non-zero (as resizing operation involves interpolation and demosaicking). In \cite{jessica:crop}, the authors proposed search for possible resizing factors using the formula $\frac{1}{1+0.005i}$ for the search range of $[max(m/p, n/q), 1]$. where $i=0,1,2, \dots R$ for $R$ that satisfies $\frac{1}{1+0.005R} \approx max(m/p, n/q)$. Then, they iteratively resize the camera fingerprint with the values of $i = (R-3, R-2, R-1, 0, 1, \dots R-4)$ and then correlate it with the query image FE.}
% sensor/image:16x12 
% video: 12x9 (video is resized 3/4) 

% now lets say image is 12x12 search range [1, max(m/p, n/q)] = [1,1]
% normally it should have been [1,3/4] with [1, min(m/p, n/q)] = [1,3/4]

% since a part of the camera sensor is used to capture an image, using the same equation may result in a narrower range. Hence, we update the above equation as follows, $\bigl[\frac{1}{r}, min(M/m, N/n)\bigr]$.
%In \cite{jessica:crop}, exhaustive search tetchiness have been used to find the in-camera resizing and cropping ratios of images. %This paper also uses the exhaustive search technique to obtain those. 

% Unlike \cite{jessica:crop}, the crop location, however, is not found using exhaustive search method. Instead a smarter search method based on experimental observations is used. 

%Another important thing to note is the fact that there might be multiple resizing operations such as capturing lower resolution image with zoom. In this case, after taking all resizing factors into consideration, the same technique can be applied for cropping ratio calculation.

% \textit{Case (iii):} In this case, we first find the in-camera resizing factor $r$ and then fix the range $[1/r, min(M/m, N/n)/r]$. While searching for the resizing factor, we also deal with orientation issue. If neither of the image and video is a square, we try two possible orientations; otherwise, we try all four possible orientations until we find a match.

\paragraph{Cropping ratio}
Although finding the correct search range is important, choosing a wide range which includes improbable resizing factors will increase time complexity. Therefore, we show a way to limit the upper bound of the search range. As described in Section~\ref{background:sensor_resizing}, excessive cropping during video capture is unlikely due to the severe impact it can have on the field of view. This provides a potential way to decrease the upper bound of the search range as explained below. 

We define \textit{cropping ratio} to be the minimum of the ratio of the number of rows and columns in the resized video frame without cropping, divided by number of rows and columns in the final video frame after cropping. 
For example, suppose a camera with a sensor resolution $4000\times 3000$ captures an HD video ($1280\times 720$). In this case, the camera might resize the raw sensor output by a factor between $1$ (i.e., no resizing, only cropping) to $\frac{1280}{4000}$ ($=0.32$, i.e., no cropping, only resizing). After resizing, the video frame can be obtained by cropping the center of the scaled region if it is still larger. With this example, the resizing factor must be between $0.32$ and $1$, and the cropping ratio is between $1$ and $3.125$. Now suppose the in-camera resizing factor is $0.5$ (i.e., output image after resizing is $2000\times 1500$), and the resized image is cropped from the center to obtain a video frame. Here, the cropping ratio is $min(\frac{2000}{1280}, \frac{1500}{720}) = 1.5625$. 

Since the order of cropping and resizing has no impact on the search range, for the sake of consistency in computation, we assume that the image is first resized and then cropped.

% PUT THE DISTRIBUTION ETC HERE
To better understand typical cropping ratios employed in cameras, we studied the distribution of cropping ratios of videos in NYUAD-mmd as shown in Fig.~\ref{fig:rsz_dist}. The figure shows that cameras tend to avoid high cropping ratios (i.e., all less than $1.6$), and most of them use a factor close to $1$ (i.e., very little or no cropping). A values less than $1$ indicates the video is captured with boundary pixels as shown in Figure~\ref{fig:rsz_dist}.

% \vspace{-1mm}
\begin{figure}[ht!]
\centering
 \includegraphics[width=74mm]{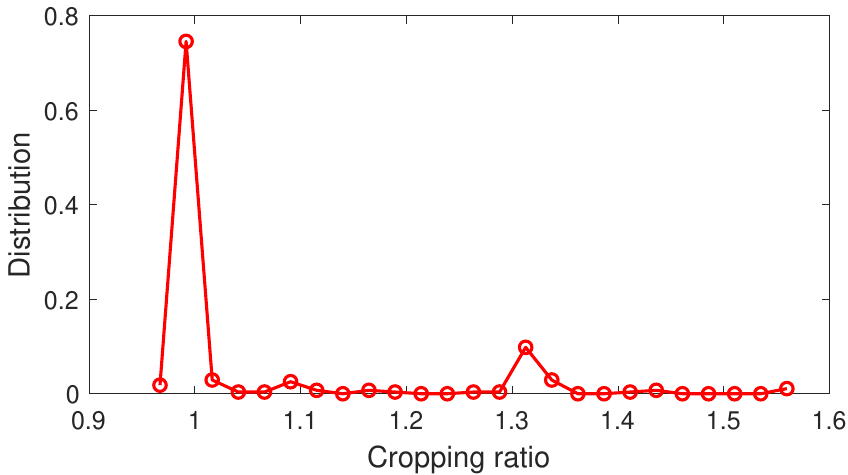}
 \caption{Distribution of cropping ratios in NYUAD-mmd}
\label{fig:rsz_dist}
\end{figure}

From these observations, we can derive two heuristics to speed up the search process. %The first one speeds up all the correlations (both matching, $H_1$, and mismatching cases, $H_0$) and the second one is specifically for matching cases. 
The first heuristic is to narrow the search space by stopping when the cropping ratio is above $1.6$. The second heuristic starts from the most likely resizing factors and progresses to lesser likely ones. 

Suppose the same camera sensor in the above example captures an HD video (which may or may not have used boundary pixels). Using the algorithm (in Fig.~\ref{fig:exh_search}), the row range of $1250$ to $4000$ (i.e., a total of $426$ correlations) is searched for finding the correct resizing factor using approach provided in \cite{jessica:crop}. In other words, the range possible cropping ratios is between $4000/1280 = 3.125$ and $1250/1280 = 0.976$. Using the knowledge, obtained from Fig.~\ref{fig:rsz_dist}, we can narrow the cropping ratio range from $0.976$ to $1.6$ as the maximum cropping ratio in the dataset is less than $1.6$. Hence the row search range can be decreased from $1250-4000$ to $1250-2048$ (i.e., a total of $235$ correlations) which can speed up close to five times. 

The second heuristic that can be derived from Fig.~\ref{fig:rsz_dist} is that the majority of cameras (i.e., $82\%$) capture videos with almost no cropping (i.e., a cropping ratio in $[0.97$, $1.1]$). When the search starts from higher cropping ratio to lower (i.e., lower resizing factors to higher in bilinear scaling), in $82\%$ of the cases a match will occur within a short time. However, this heuristic will not be useful for $H_0$ cases as there will be no match in the search. Therefore, the search will continue until trying all possible cropping/resizing factors in the range.

% Therefore, the final search range from case (i) becomes $\bigl[1/{r_1}, max(m/p, n/q)\bigr]$

\begin{table}[!h]
\centering
{\color{black}
\caption{{\color{black}Required time for search in sample $H_0$ cases}}

\begin{tabular}{ |c|c|c|c|c|c|} 
 \hline 
 image         & video      &method & max cf & $\#$ ncc & time (s) \\ \hline
\multirow{6}{*}{\begin{sideways} $4000\times3000$\end{sideways} } & \multirow{2}{*}{$1280\times720$} 
 & smart          & 1.600 & 273 & 102 \\
 & & \cite{jessica:crop}    & 3.125 & 426 & 482 \\ \cline{2-6}
 & \multirow{2}{*}{$1920 \times 1080$} & smart & 1.600 & 173 & 171 \\ 
 & & \cite{jessica:crop} & 2.083 & 218 & 247\\ \cline{2-6}
 & \multirow{2}{*}{$2560 \times 1440$} & smart & 1.563 & 123 & 198 \\ 
 & & \cite{jessica:crop} & 1.563 & 114 & 192 \\ 
\hline

\end{tabular}
\label{tbl:search_time_h0}
}
% \vspace{-1em}
\end{table}

To better understand the effect of narrowing the cropping ratio range, we compared the running time of smart search with the standard method in \cite{jessica:crop}. 
Table~\ref{tbl:search_time_h0} shows the running times (using one resizing technique) of $3$ cases where the image is $4000\times 3000$ and the video is HD ($1280\times720$), FHD ($1920\times1080$) or QHD $2560\times1440$. For these cases, we estimated maximum cropping ratio (i.e., max cf), number of NCC operations ($\#$ ncc) and total time in seconds for smart and exhaustive search.

%against the exhaustive search \cite{jessica:crop} with two media from random cameras. For three sample cases, we compared the time required for both search techniques. 
The results show that when the image resolution ratio of the image and video is high, the speedup achieved by smart search is significant. When the difference is low, smart search can be slightly slower as the number of NCC computations increase since we crop the boundary pixels.
}

{\color{black}
It should be noted that once the resizing factor for a camera model is found, we don't need to calculate it again. We can create a lookup table which contains the resizing factor (or the matching resolution) corresponding to each possible pair of media objects from a particular camera model. 
Table~\ref{tbl:lut} shows this information for the Xiaomi Redmi Note 3 and Nexus 5 smartphone cameras.
For example, an image with $4160\times3120$ can be matched with a Full HD video when it is resized with a factor of $0.4563$ such that its final resolution becomes $1898\times1424$. This way, the true peak can be found using NCC, and the video and the image can be matched. It is crucial to note that because in-camera resizing is software dependent, cameras of a particular model with different software may have different resizing factors; however, we haven't observed such a case in NYUAD-mmd.}
%Along with these, it contains the resizing factor (rf in the table) of the images which lead to a match with video and final resolution of the image after resizing. 

\begin{table}[!h]
\centering
{\color{black}
\caption{{\color{black}Parameter look up table for sample cameras. ``Match resol." stands for the matching resolution that image is resized to, and ``rf" indicates the resizing factor in this case.}}

\begin{tabular}{ |c|c|c|c|c|c|} 
 \hline 
camera & image         & video      & match resol & rf \\ \hline
\multirow{2}{*}{Redmi N3} & $4160\times3120$ & $1280\times720$ & $1263\times948$ & 0.3036 \\
 & $4160\times3120$ & $1920\times1080$    & $1898\times1424$ & 0.4563\\ \hline
\multirow{3}{*}{Nexus 5} & $3264\times2448$ & $1920\times1080$ & $1920\times1440$ & 0.5882\\
&$3264\times2448$ & $1280\times720$ & $1280\times960$ & 0.3922\\
&$3200\times2368$ & $1920\times1080$ & $1883\times1394$ & 0.5884\\
\hline
\end{tabular}
\label{tbl:lut}
}
% \vspace{-1em}
\end{table}

\vspace{-4mm}
\section{Experimental Analysis}
\label{sec:experiment}
For evaluation, we created a dataset, NYUAD mixed media dataset (\textit{NYUAD-mmd}), which contains visual media from $78$ smartphone cameras ($19$ brands, $62$ models). From these cameras, a total of $6892$ images, and $301$ non-stabilized videos (most of them being 40+ seconds) of different resolutions, as allowable by the camera settings, were collected. Most cameras allow images or video to be captured at more than one resolution. For each camera, images with the same resolution were grouped together to calculate a fingerprint, corresponding to that resolution. Next, we used the first 40 seconds (i.e., approx. $1200$ frames) of each video to create a video FE. NCC was used to determine potential alignments between two fingerprints and PCE for testing if they do indeed match. The performance (i.e., true positive rate and false positive rate) of the proposed method was compared with {\color{black}\cite{jessica:crop} and} our previous work \cite{taspinar2016source}.
All the experiments were implemented on Matlab 2016a on Windows 7 PC with 32 GB memory and Intel Xeon(R) E5-2687W v2 @3.40GHz CPU. 

% \vspace{-3mm}
%\subsection{Mixed media}
% \vspace{-2mm}
\paragraph{Experiment: Train on images, test on videos}
For this experiment, a total of $149$ image FEs were computed from $62$ cameras as the dataset contains at least $2$ different image resolutions for most cameras. A video FE was extracted from each video. So a total of $301$ video FEs were computed. 

{\color{black}Since both image FEs and video FEs are reliable (from many still images), based on the analysis in Section~\ref{sec:math} and the algorithm suggestion in Section \ref{sec:approach}, bilinear scaling was the only resizing technique tried. %that all the resizing operations for this experiment is done using bilinear scaling; we did not use binning or line-skipping for the resizing operations in smart search. This is because hence it is unlikely to achieve a better performance using binning for this particular experiment. 
 }
 
We correlated all image FEs with the video FEs taken by the same camera (i.e., alternate hypothesis, $H_1$). We then compared each image FE with random video FEs to evaluate FPR (\textit{i.e.} null hypothesis, $H_0$).
A total of $583$ correlations were performed for $H_1$ and $563$ for $H_0$ by comparing image FEs of $i^{\rm th}$ with the video FEs of $(i+1)^{\rm th}$ camera. In this experiment, we compared our results with {\color{black} \cite{jessica:crop} and} \cite{taspinar2016source}. 

% Table~\ref{tbl:imFP_vidFP} shows the performance of the proposed method and the method proposed in \cite{taspinar2016source}.

% As shown in the table, the TPR of the proposed method is approximately $10.8\%$ higher than the method in \cite{taspinar2016source}.
%This is because the proposed method is designed by keeping in mind that boundary pixels are used to take a video. In \cite{taspinar2016source}, this fact was ignored. On the other hand, the FPR is the same in both cases (\textit{i.e.,} $1.07\%$). 

{\color{black}
The results show that when $\tau$ is set to $60$, TPR is $65.52\%$, $86.11\%$ and $96.91\%$ for \cite{jessica:crop}, \cite{taspinar2016source} and the proposed method, respectively. The main reason for \cite{jessica:crop} to perform significantly lower is because it doesn't consider that the images taken with lower resolution might be scaled and cropped.
Since smartphones were not commonly used at the time of publication of \cite{jessica:crop}, the problems this paper addresses may not have been relevant then.} The $10\%$ lower TPR for \cite{taspinar2016source} is due to the boundary pixel issue as well as failing to see the lower resolution images might be already cropped and resized (within the camera) which happens in $24$ cases. Further details for the cameras (use of boundary pixels etc.) can be found in Appendix~\ref{appendix:cameras}.

\begin{table}[!h]
\centering
\caption{Performance of train on images, test on videos}

{\color{black}
\begin{tabular}{ |c|c|c|c|c|c|c|} 
 \hline 
   & method& $\#match$ & $total$ & $TPR$ & $FPR$ & Time(sec)\\ \hline
    
 \multirow{3}{*}{$H_0$} & \cite{jessica:crop} & 4 & 563 & N/A & 0.71\% &469 \\
   & \cite{taspinar2016source} & 6 & 563 & N/A & 1.07\% & 567 \\
   & smart & 6 & 563 & N/A & 1.07\% & 148 \\ \hline
 \multirow{3}{*}{$H_1$} & \cite{jessica:crop} & 382 & 583 & 65.52\% & N/A\% & 171\\
   & \cite{taspinar2016source} & 502 & 583 & 86.11\% & N/A\% & 93\\
    % & \cite{taspinar2016source} & 444 & 583 & 76.16\% & N/A\%\\
   & smart & 565 & 583 & 96.91\% & N/A\% & 47 \\ \hline
\end{tabular}
}
\label{tbl:imFP_vidFP}
% \vspace{-1em}
\end{table}

{\color{black} Note that we used Matlab's bilinear scaling rather than the bilinear scaling used for analysis in Section~\ref{sec:math}. When the same experiment is done with the filter used in Section~\ref{sec:math}, the TPR drops down from $96.91\%$ to $\frac{557}{583}=95.54\%$ and average PCE drops from $17144$ to $16325$.}

%Further examination of the results on a per camera basis and a comparison to \cite{taspinar2016source} showed that Lenovo P1, Lenovo S90, Motorola G3, Xiaomi Redmi Note3, Nexus 6, Oppo A57, XOLO Black-1X are some of the cameras that use boundary pixels while capturing videos but not while capturing images. Hence \cite{taspinar2016source} had failed to make correct attribution with them. Appendix A provides a more comprehensive list of such insights obtained by our experiments. 

% \vspace{-1mm}
\paragraph{Experiment: Train on videos, test on images}
% \label{experiment:vF2iN}
In this experiment, each image FE (here image FE refers to the PRNU noise of a single image) taken by a specific camera was correlated with all video FEs of the same camera. Since the image FE estimate is not highly accurate, based on the analysis in Section~\ref{sec:math} and the algorithm suggestion in Section \ref{sec:approach}, both bilinear scaling and binning were tried for resizing. As we know from RoA derivation in Section~\ref{sec:math}, bilinear scaling may not be the best approach to resize the images when the video is resized with binning especially when FE quality is low. To resize using binning,  we reverted RGB images to raw and using $2\times2$ binning scheme, we downsized the raw image to half resolution. We then followed the basic imaging pipeline and using demosaicing and bilinear scaling, and obtained an image FE which is at the resolution that we already learnt from the previous experiment.

%This approach is similar to how cameras downsize videos when they use binning. We expect this scheme to perform better than bilinear scaling for the cases when the video is resized with binning (i.e., for $2\times2$ binning).

{\color{black} In our experiment, we first compared all video FEs with image FEs using bilinear scaling for both \cite{taspinar2016source} and smart search. The TPRs for these two cases were $74.55\%$ and $82.85\%$ respectively, as shown in Table~\ref{tbl:vidFP_imNS}. 
%In such cases, binning can be more useful. To evaluate the correctness of this idea, we reverted RGB images to raw. Then, resized the raw images to half resolution with binning and then downsized them using bilinear scaling to the resolutions we already learnt from the previous experiment. Thus, in a way we emulated binning.
% THE FOLLOWING TEST IS NOT CLEAR. I COULD NOT UNDERSTAND. CAN YOU PLEASE SIMPLIFY.
% and the image didn't go through in-camera downsizing step.%, however, when there is in-camera resizing applied on images this scheme is unlikely to work even will have lower PCE result. 
Then we used resizing using only binning and achieved a TPR of $79.84\%$ (labeled \textit{bin} in the table). Finally, we get the maximum PCE values of binning and bilinear scaling which resulted in $85.72\%$ TPR. These results show that in $3\%$ of all cases (i.e., $692$ of $24077$ cases) images don't match with the corresponding video FE when they are downsized with bilinear scaling whereas they match with binning. These results show even if we know the correct resizing factors, we may not be able to match due to differences in video and image resizing techniques (i.e., due to lower RoA). However, when we resize with a technique that potentially has higher RoA, we may be able to match.}

{\color{black} Note that since \cite{jessica:crop} was proposed as a general solution for cropping and resizing, it did not capture some of the cases in this experiment. Hence, we compare the proposed approach with only \cite{taspinar2016source}}.

% As shown in Table~\ref{tbl:vidFP_imNS}, the TPR of the proposed scheme was approx. $8\%$ higher than the scheme proposed in \cite{taspinar2016source}. For both the schemes, the TPR is lower for the PRNU noise of a single image than when the fingerprint computed from an image set is used (Table~\ref{tbl:imFP_vidFP}) as the PRNU noise obtained from a single image can contain other noise whereas when multiple images are used, other noise patterns can be suppressed\cite{fridrich2013sensor}.

% Note that when two image FEs with different resolutions are present, choosing the higher resolution fingerprint is typically more reliable when they are correlated with a video FE. This is because the higher resolution image might have been taken using the full camera sensor whereas the smaller one might have been captured by some part of the camera (i.e., the center of the camera might have been used). This might cause a problem for determining the search range of the resizing factor. We observed this issue in $25$ of the $583$ experimental cases. 

\begin{table}[!ht]
 \centering
 \caption{{\color{black}Performance of train on videos, test on images}}
 
 {\color{black}\begin{tabular}{ |c|c|c|c|c| } \hline
 type  & $\#match$ & $\#comparison$ & $ TPR $ \\ \hline
 \cite{taspinar2016source} & 17949 & 24077 & 74.55\% \\ \hline
 bilinear & 19948 & 24077 & 82.85\% \\ \hline
 bin  & 19224 & 24077 & 79.84\% \\ \hline
 bin+bilinear & 20640 & 24077 & 85.72\% \\ \hline
 \end{tabular}}
 \label{tbl:vidFP_imNS}
 \vspace{-0.3em}
\end{table}

\begin{figure}[t!]
\includegraphics[width=83mm]{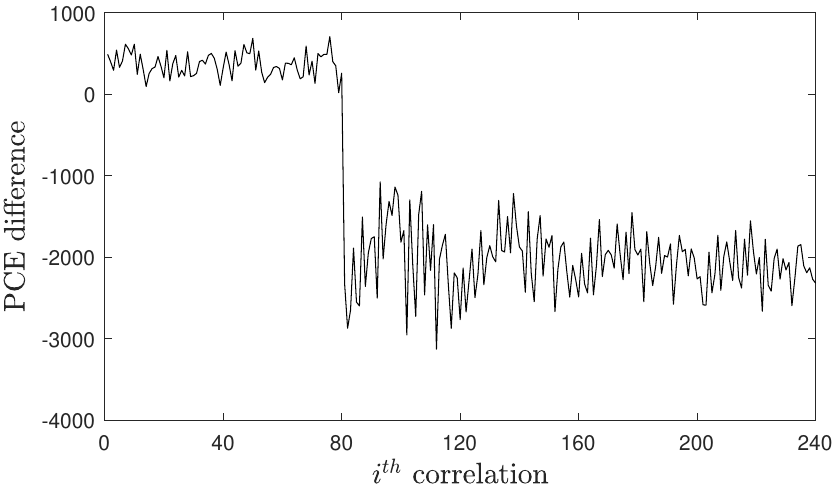}
\caption{{\color{black}PCE differences of video-image comparisons of a Xiaomi Note $4$ camera when images are resizing with bilinear scaling and binning. The first 80 ( $1-80^{th}$) correlations are with HD videos and the rest ($81-240^{th}$) with Full HD videos.}}
 % \vspace{-1mm}
\label{fig:xiaomi}
\end{figure}

{\color{black}
Analysis of our experimental results also revealed that the same camera might use different resizing techniques when capturing different resolution media. For example, Figure~\ref{fig:xiaomi} shows the correlations of video FEs with image FEs for a Xiaomi Note 4 camera. Two HD videos ($1280\times720$) and four Full HD videos ($1920\times1080$) and $40$ images of resolutions $4160\times3120$ were captured by the camera. We first resized the images with bilinear scaling and binning (as explained in Section~\ref{sec:math}). Then, we found the PCE (using NCC and getting the peak position after resizing) for both cases and got their differences. As seen in the figure, when the images are resized with binning, for HD videos, the PCE is higher whereas for Full HD videos, resizing with binning significantly drops the PCE value. Our inference from these results is that the HD videos may be resized with binning. Thus, they match better when images are also resized with binning. For full HD videos, binning might not be used, so resizing with binning performed poorly. Similar behavior was observed in the other two Xiaomi Note 4 cameras in the dataset.}

\paragraph{DARPA Medifor Camera ID Challenge}

In July 2018, DARPA's Media Forensics (MediFor) program conducted a PRNU-based camera attribution challenge which consisted of $6$ sub-challenges based on the type of training and test sets. For example, if we use videos and images for training, and only videos for testing, the sub-challenge was called \textit{train on multimedia, test on video}.

Participants had two options for each verification task: (i) submit an answer by providing ``confidence score", which indicates how likely the specified camera has taken the probe media, or (ii) opting out from submitting a solution (i.e., when the participant is not comfortable with the confidence score). Submissions were evaluated using three metrics: Area Under Curve (AUC), Correct Detection at $0.05$ False Acceptance Rate (CD$@0.05$ FAR) and Trial Response Rate (TRR). AUC is the area under the curve for the ROC curve that is obtained from the confidence scores. This has a value between zero and one where one indicates the perfect result.
CD$@0.05$ FAR shows how many true cases are accepted when only $5\%$ of the false cases are accepted as true. Finally, TRR indicates the rate of the tasks that were opted in.

Table~\ref{tab:darpa:fullset} shows the results for each sub-challenge when all tasks were opted in. For brevity, the name of the sub-challenges is shortened in the table. For example, ``image-mm" indicates ``test on image, train on multimedia", ``$\#P$" shows the number of participants in a sub-challenge, ``Rank" is our ranking in terms of AUC, and ``AUC+" and ``CD+" shows the difference we have with highest performing team in terms of AUC and CD, respectively. For the sub-challenges that no other group participated, we considered AUC+ and CD+ as N/A. As shown in Table~\ref{tab:darpa:fullset}, in four of the challenges that contained mixed media, no other group had submissions. Our submissions for these challenges were based on the methods proposed in this paper.

% \vspace{-1mm}
\begin{table}[!h]
 \centering
 \def\arraystretch{1.15}
 \caption{Performance comparison in fullset}
 
 \begin{tabular}{|l|c|c|cc|cc|}
 \hline
 Challenge & \#P & Rank & AU & CD & AUC+ & CD+ \\ \hline
 image$-$image & 9 & 1 & 0.87 & 0.77 & 0.07 & 0.20 \\ \hline
 image$-$mm & 1 & 1 & 0.84 & 0.69 & N/A & N/A \\ \hline
 image$-$video & 1 & 1 & 0.62 & 0.40 & N/A & N/A \\ \hline
 video$-$image & 1 & 1 & 0.76 & 0.50 & N/A & N/A \\ \hline
 video$-$mm & 1 & 1 & 0.68 & 0.47 & N/A & N/A \\ \hline
 video$-$video & 4 & 2 & 0.60 & 0.36 & -0.10 & 0.02 \\ \hline
 \end{tabular}
 \label{tab:darpa:fullset}
\end{table}

Table~\ref{tab:darpa:subset} shows the results for three sub-challenges when there was an option of opting out from submitting a solution. 

% \vspace{-1mm}
\begin{table}[!h]
 \centering
 \def\arraystretch{1.15}
 
 \caption{Performance comparison in subset}
 \begin{tabular}{|l|c|c|cc|cc|c|}
 \hline
 Challenge & \#P & Rank & AUC & CD & AUC+ & CD+ & TRR \\ \hline
 image\text{-}image & 6 & 1 & 0.99 & 0.99 & 0.05 & 0.15 & 0.58 \\ \hline
 image\text{-}video & 1 & 1 & 0.90 & 0.81 & N/A & N/A & 0.40 \\ \hline
 video\text{-}video & 4 & 2 & 0.76 & 0.58 & -0.10 & -0.10 & 0.62 \\ \hline
 \end{tabular}
 \label{tab:darpa:subset}
\end{table}

Note that since the DARPA dataset contained lower quality images and videos (e.g., stabilized, low intensity, scaled and/or cropped, tampered and so on), the error rates using this dataset is higher than NYUAD-mmd where images and video were not processed in any way after capture.

\vspace{-4mm}
\section{Conclusion and Future Work}
\label{sec:conclusion}
%PRNU-based source camera attribution method can become ineffective when the visual media objects have been captured differently by the source camera. This paper examines how a camera captures a still image. Then, we illustrate how the verification of two visual media can be achieved. The paper considers many possible use cases including when the two media have different resolutions and/or aspect ratios. The results show that this work is superior to the previous research concerning performance/TPR and comprehensiveness (i.e., covering various aspects of the problem).

PRNU-based source camera attribution may become ineffective when the reference and query media are of different types (i.e., one video and the other image). This is due to the misalignment caused by the differences between in- and out-camera operations applied on the two media. In this paper, we examined these differences and proposed the notion of ``Ratio of Alignment", RoA, which provides {\color{black}insight about how the correlation of two media will be affected due to desynchronization caused by different resizing approaches}
%a rough upper limit for correlation of the media that are desynchronized. 
We validated this analytical RoA estimation with an experiment. 

We then presented an approach for source attribution for mixed media based on the knowledge obtained for in-camera processing and RoA analysis. The approach was validated using experiments on a dataset consisting of mixed media (i.e., reference is a set of images and query is a video or reference is a video and query is a single image). It was shown that the proposed approach gives state-of-the-art results. Although experimental results using our dataset involved pristine media (i.e., not modified outside the camera), experiments with a DARPA dataset that included modified images and video, resulted in a good performance overall as well. Our experiments also revealed insights about in-camera processing for different camera models as listed in Appendix A. %~\ref{sec:appendix}.  

One of the biggest challenges while performing source camera attribution is the development of efficient and effective techniques to determine resizing factors. %This is due to the fact that current techniques. 
Since reverse engineering the resizing factor of a media is often infeasible, it is crucial to come up with techniques that will compute the resizing technique and factor in an efficient manner rather than trying all possible parameters. Although there has been some effort for resizing factor determination techniques for images, there is significant room, and they typically don't do well for videos based on our experience. There has been no work in resizing technique determination. In fact, media forensics research has not taken into account the fact that different resizing techniques can be deployed as has been done in this work. 

Another avenue for future research stems from the fact that RoA estimation showed here has a significant performance drop caused by different resizing techniques. Perhaps one can develop an out-camera resizing technique that has higher RoA and achieve higher correlation with videos resized by commonly used techniques such as binning and line-skipping. 

Besides, when the reference is a video, and the query is images, the attribution performance drops to $82.85\%$ in the best case. Clearly, there is a need for improvement. For example, a technique to obtain a better quality PRNU noise from videos may help achieve better performance. 

Finally,  we have assumed that neither the video nor the images are zoomed, stabilized,  or obtained using a non-linear operation (such as HDR). Moreover, we think they were not subjected to an out-camera cropping and/or resizing operation. Although there has been research lately to perform attribution in the presence of such operations, a lot more work is needed to obtain higher accuracy.  Although such techniques would supplement the work presented in these papers, the attribution performance drop when a cascade of such techniques is applied need to be determined. 

%when the visual media objects have been captured differently by the source camera. 
%This paper examines how a camera captures a still image. Then, we illustrate how the verification of two visual media can be achieved. The paper considers many possible use cases including when the two media have different resolutions and/or aspect ratios. The results show that this work is superior to the previous research concerning performance/TPR and comprehensiveness (i.e., covering various aspects of the problem).

%\section{Acknowledgement}
%We thank Dr. Mehmet Aktas and Dr. Fatih Erden for their valuable contributions in the theoretical analysis which greatly improve the manuscript. %and validating the correctness of them...

% Maybe the students who collect the visual media

\vspace{-3mm}
\bibliographystyle{IEEEtran}
\bibliography{main.bbl}

\begin{thebibliography}{10}
\providecommand{\url}[1]{#1}
\csname url@samestyle\endcsname
\providecommand{\newblock}{\relax}
\providecommand{\bibinfo}[2]{#2}
\providecommand{\BIBentrySTDinterwordspacing}{\spaceskip=0pt\relax}
\providecommand{\BIBentryALTinterwordstretchfactor}{4}
\providecommand{\BIBentryALTinterwordspacing}{\spaceskip=\fontdimen2\font plus
\BIBentryALTinterwordstretchfactor\fontdimen3\font minus
  \fontdimen4\font\relax}
\providecommand{\BIBforeignlanguage}[2]{{%
\expandafter\ifx\csname l@#1\endcsname\relax
\typeout{** WARNING: IEEEtran.bst: No hyphenation pattern has been}%
\typeout{** loaded for the language `#1'. Using the pattern for}%
\typeout{** the default language instead.}%
\else
\language=\csname l@#1\endcsname
\fi
#2}}
\providecommand{\BIBdecl}{\relax}
\BIBdecl

\bibitem{TahaBook2013}
H.~T. Sencar and N.~Memon, \emph{Digital image forensics: There is more to a
  picture than meets the eye}.\hskip 1em plus 0.5em minus 0.4em\relax New York,
  USA: Springer, 2013.

\bibitem{fridrich2009digital}
J.~Fridrich, ``Digital image forensics,'' \emph{IEEE Signal Processing
  Magazine}, vol.~26, no.~2, 2009.

\bibitem{delp2009digital}
E.~Delp, N.~Memon, and M.~Wu, ``Digital forensics,'' \emph{IEEE Signal
  Processing Magazine}, vol.~26, no.~2, pp. 14--15, 2009.

\bibitem{milani2012overview}
S.~Milani, M.~Fontani, P.~Bestagini, M.~Barni, A.~Piva, M.~Tagliasacchi, and
  S.~Tubaro, ``An overview on video forensics,'' \emph{APSIPA Transactions on
  Signal and Information Processing}, vol.~1, 2012.

\bibitem{lukas2006digital}
J.~Lukas, J.~Fridrich, and M.~Goljan, ``Digital camera identification from
  sensor pattern noise,'' \emph{IEEE TIFS}, vol.~1, no.~2, pp. 205--214, 2006.

\bibitem{jessica:crop}
M.~Goljan and J.~Fridrich, ``Camera identification from scaled and cropped
  images,'' \emph{Proc. SPIE, Electronic Imaging, Forensics, Security,
  Steganography, and Watermarking of Multimedia Contents X}, vol. 6819, pp.
  68\,190E--68\,190E--13, 2008.

\bibitem{yaqub2018towards}
W.~Yaqub, M.~Mohanty, and N.~Memon, ``Towards camera identification from
  cropped query images,'' in \emph{25th ICIP}.\hskip 1em plus 0.5em minus
  0.4em\relax IEEE, 2018, pp. 3798--3802.

\bibitem{bayram2013seam}
S.~Bayram, H.~Sencar, and N.~Memon, ``Seam-carving based anonymization against
  image \& video source attribution,'' in \emph{MMSP, 2013 IEEE 15th
  International Workshop on}.\hskip 1em plus 0.5em minus 0.4em\relax IEEE,
  2013, pp. 272--277.

\bibitem{dirik2014analysis}
A.~E. Dirik, H.~T. Sencar, and N.~Memon, ``Analysis of seam-carving-based
  anonymization of images against prnu noise pattern-based source
  attribution,'' \emph{IEEE TIFS}, vol.~9, no.~12, pp. 2277--2290, 2014.

\bibitem{taspinar2017prnu}
S.~Taspinar, M.~Mohanty, and N.~Memon, ``Prnu-based camera attribution from
  multiple seam-carved images,'' \emph{IEEE TIFS}, vol.~12, no.~12, pp.
  3065--3080, 2017.

\bibitem{taspinar2016prnu}
------, ``Prnu based source attribution with a collection of seam-carved
  images,'' in \emph{2016 IEEE ICIP}.\hskip 1em plus 0.5em minus 0.4em\relax
  IEEE, 2016, pp. 156--160.

\bibitem{mandelli2017inpainting}
S.~Mandelli, L.~Bondi, S.~Lameri, V.~Lipari, P.~Bestagini, and S.~Tubaro,
  ``Inpainting-based camera anonymization,'' in \emph{Image Processing (ICIP),
  2017 IEEE International Conference on}.\hskip 1em plus 0.5em minus
  0.4em\relax IEEE, 2017, pp. 1522--1526.

\bibitem{John:2016:Patch}
J.~Entrieri and M.~Kirchner, ``Patch-based desynchronization of digital camera
  sensor fingerprints,'' in \emph{IS\&T Media Watermarking, Security, and
  Forensics}, 2016.

\bibitem{Karak:2015:PRNU}
A.~Karak\"{u}\c{c}\"{u}k, A.~E. Dirik, H.~T. Sencar, and N.~Memon, ``Recent
  advances in counter {PRNU} based source attribution and beyond,'' \emph{IS\&T
  Media Watermarking, Security, and Forensics}, vol. 9409, April 2015.

\bibitem{darpa2018medifor}
``{DARPA Medifor Program},''
  \url{https://www.darpa.mil/program/media-forensics/}, 2018, [Online; accessed
  7-August-2018].

\bibitem{bayer1976color}
B.~E. Bayer, ``Color imaging array,'' Jul.~20 1976, uS Patent 3,971,065.

\bibitem{jin2012analysis}
X.~Jin and K.~Hirakawa, ``Analysis and processing of pixel binning for color
  image sensor,'' \emph{EURASIP Journal on Advances in Signal Processing}, vol.
  2012, no.~1, p. 125, 2012.

\bibitem{zhang2018pixel}
J.~Zhang, J.~Jia, A.~Sheng, and K.~Hirakawa, ``Pixel binning for high dynamic
  range color image sensor using square sampling lattice,'' \emph{IEEE
  Transactions on Image Processing}, vol.~27, no.~5, pp. 2229--2241, 2018.

\bibitem{taspinar2016source}
S.~Taspinar, M.~Mohanty, and N.~Memon, ``Source camera attribution using
  stabilized video,'' in \emph{2016 IEEE International Workshop on Information
  Forensics and Security (WIFS)}.\hskip 1em plus 0.5em minus 0.4em\relax IEEE,
  2016, pp. 1--6.

\bibitem{imaging2004mt9m001}
A.~Imaging, ``Mt9m001: 1/2-inch megapixel cmos digital image sensor,''
  \emph{MT9M001 DS Rev}, vol.~1, pp. 1--27, 2004.

\bibitem{mihcak1999low}
M.~K. Mihcak, I.~Kozintsev, K.~Ramchandran, and P.~Moulin, ``Low-complexity
  image denoising based on statistical modeling of wavelet coefficients,''
  \emph{IEEE Signal Processing Letters}, vol.~6, no.~12, pp. 300--303, 1999.

\bibitem{dabov2009bm3d}
K.~Dabov, A.~Foi, V.~Katkovnik, and K.~Egiazarian, ``Bm3d image denoising with
  shape-adaptive principal component analysis,'' in \emph{SPARS'09-Signal
  Processing with Adaptive Sparse Structured Representations}, 2009.

\bibitem{lewis1995fast}
J.~P. Lewis, ``Fast normalized cross-correlation,'' in \emph{Vision interface},
  vol.~10, no.~1, 1995, pp. 120--123.

\bibitem{jessica:camera}
J.~Luk\'{a}\v{s}, J.~Fridrich, and M.~Goljan, ``Digital camera identification
  from sensor pattern noise,'' \emph{IEEE TIFS}, vol.~1, no.~2, pp. 205--214,
  2006.

\bibitem{Sutcu:2007:Improve}
Y.~Sutcu, S.~Bayram, H.~T. Sencar, and N.~Memon, ``Improvements on sensor noise
  based source camera identification,'' in \emph{IEEE International Conference
  on Multimedia and Expo}, 2007, pp. 24--27.

\bibitem{Li:2012:Color}
C.~T. Li and Y.~Li, ``Color-decoupled photo response non-uniformity for digital
  image forensics,'' \emph{IEEE Transactions on Circuits and Systems for Video
  Technology}, vol.~22, no.~2, pp. 260--271, 2012.

\bibitem{Chierchia:2010:IDP}
G.~Chierchia, S.~Parrilli, G.~Poggi, C.~Sansone, and L.~Verdoliva, ``On the
  influence of denoising in {PRNU} based forgery detection,'' in \emph{ACM
  Multimedia in Forensics, Security and Intelligence}, 2010, pp. 117--122.

\bibitem{Li:2010:Enhance}
C.~T. Li, ``Source camera identification using enhanced sensor pattern noise,''
  \emph{IEEE TIFS}, vol.~5, no.~2, pp. 280--287, 2010.

\bibitem{Caldelli:201:WIFS}
R.~Caldelli, I.~Amerini, F.~Picchioni, and M.~Innocenti, ``Fast image
  clustering of unknown source images,'' in \emph{IEEE International WIFS},
  2010, pp. 1--5.

\bibitem{lukavs2006detecting}
J.~Luk{\'a}{\v{s}}, J.~Fridrich, and M.~Goljan, ``Detecting digital image
  forgeries using sensor pattern noise,'' in \emph{Electronic Imaging
  2006}.\hskip 1em plus 0.5em minus 0.4em\relax International Society for
  Optics and Photonics, 2006, pp. 60\,720Y--60\,720Y.

\bibitem{Chierchia:2014:Fing}
G.~Chierchia, G.~Poggi, C.~Sansone, and L.~Verdoliva, ``A {Bayesian-MRF}
  approach for {PRNU}-based image forgery detection,'' \emph{IEEE TIFS},
  vol.~9, no.~4, pp. 554--567, 2014.

\bibitem{Simone:VideoOv:2012}
S.~Milani, M.~Fontani, and P.~B. et. al., ``An overview on video forensics,''
  \emph{Signal Processing Systems}, vol.~1, pp. 1--18, June 2012.

\bibitem{Chen:2007:Vid}
M.~Chen, J.~Fridrich, M.~Goljan, and J.~Lukas, ``Source digital camcorder
  identification using sensor photo response non-uniformity,'' in \emph{SPIE
  Electronic Imaging}, 2007, pp. 1G--1H.

\bibitem{McCloskey:2008:Confidence}
S.~McCloskey, ``Confidence weighting for sensor fingerprinting,'' in \emph{IEEE
  CVPR Workshops}, 2008, pp. 1--6.

\bibitem{Chuang:2011:Vcomp}
W.-H. Chuang, H.~Su, and M.~Wu, ``Exploring compression effects for improved
  source camera identification using strongly compressed video,'' in \emph{IEEE
  ICIP}, 2011, pp. 1953--1956.

\bibitem{Chen:2013:Video}
S.~Chen, A.~Pande, K.~Zeng, and P.~Mohapatra, ``Video source identification in
  lossy wireless networks,'' in \emph{IEEE INFOCOM}, 2013, pp. 215--219.

\bibitem{Houten:2009:YVid}
W.~van Houten and Z.~Geradts, ``Using sensor noise to identify low resolution
  compressed videos from {Y}outube,'' in \emph{IAPR International Workshop on
  Computational Forensics}, 2009, pp. 104--115.

\bibitem{hyun2012camcorder}
D.-K. Hyun, C.-H. Choi, and H.-K. Lee, ``Camcorder identification for heavily
  compressed low resolution videos,'' in \emph{Computer Science and
  Convergence}.\hskip 1em plus 0.5em minus 0.4em\relax Springer, 2012, pp.
  695--701.

\bibitem{iuliani2017hybrid}
M.~Iuliani, M.~Fontani, D.~Shullani, and A.~Piva, ``A hybrid approach to video
  source identification,'' \emph{arXiv preprint arXiv:1705.01854}, 2017.

\bibitem{dang2015raise}
D.-T. Dang-Nguyen, C.~Pasquini, V.~Conotter, and G.~Boato, ``Raise: a raw
  images dataset for digital image forensics,'' in \emph{Proceedings of the 6th
  ACM Multimedia Systems Conference}.\hskip 1em plus 0.5em minus 0.4em\relax
  ACM, 2015, pp. 219--224.

\bibitem{mandelli2018facing}
S.~Mandelli, P.~Bestagini, L.~Verdoliva, and S.~Tubaro, ``Facing device
  attribution problem for stabilized video sequences,'' \emph{arXiv preprint
  arXiv:1811.01820}, 2018.

\bibitem{Miroslav:Private}
M.~Goljan, ``Camera identification from hdr images,'' Private conversation.

\bibitem{gallagher2005detection}
A.~C. Gallagher, ``Detection of linear and cubic interpolation in jpeg
  compressed images,'' in \emph{null}.\hskip 1em plus 0.5em minus 0.4em\relax
  IEEE, 2005, pp. 65--72.

\bibitem{mahdian2008blind}
B.~Mahdian and S.~Saic, ``Blind authentication using periodic properties of
  interpolation,'' \emph{IEEE Transactions on Information Forensics and
  Security}, vol.~3, no.~3, pp. 529--538, 2008.

\end{thebibliography}

\vspace{-4mm}
\begin{appendices}
\label{sec:appendix}
    
{\color{black}
\section{Mathematical Analysis}

\subsubsection{Sensor - Pixel correspondence for bilinear scaling for Green component} \label{apdx:bilinear}
For the green color plane, the $(m, n)^{\rm th}$ pixel of the demosaiced image can be obtained as
\begin{equation}
\small
    I^{temp}_G(m, n) = 
    \begin{cases}
        \frac{I(2m-2:2:2m,2n-1)+I(2m-1,2n-2:2:2n)}{4}   &\text{case 1 }\\ 
        {\scriptstyle I(2m-1,2n)  }                                    &\text{case 2 }\\ 
        {\scriptstyle I(2m,2n-1) }                                     &\text{case 3 }\\ 
        \frac{I(2m-1:2:2m+1,2n)+I(2m,2n-1:2:2n+1)}{4}   &\text{case 4.} 
    \end{cases}
    \label{eqn:cfa:green}
\end{equation}

After bilinear scaling, the green value of the $(m, n)^{\rm th}$ pixel of the scaled image, $I_G^{Bscale}(m,n)$, is given as
\begin{equation}
\small
    \begin{aligned}
    I_G^{Bscale}(m,n) = \frac{1}{4} \Big( & I^{temp}_G(2m-1,2n-1) + I^{temp}_G(2m-1,2n) + \\
                                          & I^{temp}_G(2m,2n-1) + I^{temp}_G(2m,2n)\Big).
    \end{aligned}
    \label{eqn:bilinear:green}
\end{equation}

Putting~(\ref{eqn:cfa:green}) in~(\ref{eqn:bilinear:green}), we obtain
\begin{equation}
\small
    \begin{aligned}
        I_G^{Bscale}(m,n) = &\frac{3}{8} \Big(I(2m-1,2n)+I(2m,2n-1)\Big) + \\ 
                            &\frac{1}{16} \Big(I(2m-2,2n-1)  + I(2m-1,2n-2) + \\
                            & ~~~~~~I(2m,2n+1)+I(2m+1,2n)\Big).
    \end{aligned}
    \label{eqn:bilinear:green:final}
\end{equation}

\subsubsection{Sensor-Pixel Correspondences for Binning}
\label{apdx:binning}

Assume that the $I_G^{temp}(m, n)$ represents the green color component of the $(m, n)^{\rm th}$ pixel of the binned sensor output which can be found as 
%On the other hand, the green color component for ($i, j$) pixel of binned sensor can be given as 
\begin{equation}
\small
    I_G^{temp}(m, n) = 
    \begin{cases}
        \frac{I(2m-1:2:2m+1,2n-2:2:2n)}{4}     &\text{case 2} \\
        \frac{I(2m-2:2:2m,2n-1:2:2n+1)}{4}     &\text{case 3} \\
        0                                      &\text{otherwise}.
    \end{cases}
    \label{eqn:binning:green:init}
\end{equation}

After color filter array demosaicing, the green color component for  $I_G^{Bin}(m,n)$ becomes
\begin{equation}
\small
    \begin{aligned}
    &I_G^{Bin}(m,n) =
    \begin{cases}
        \frac{I_G^{temp}(m-1:2:m+1,n)+I_G^{temp}(m,n-1:2:n+1)}{4}  &\text{case 1} \\
        {\scriptstyle I_G^{temp}(m,n) }                    &\text{case 2}\\
        {\scriptstyle I_G^{temp}(m,n) }                    &\text{case 3}\\
        \frac{I_G^{temp}(m-1:2:m+1,n)+I_G^{temp}(m,n-1:2:n+1)}{4}  &\text{case 4}.\\
    \end{cases}
    \end{aligned}
    \label{eqn:binning:green:second}
\end{equation}

Putting~(\ref{eqn:binning:green:init}) in~(\ref{eqn:binning:green:second}), we  find $I_G^{Bin}(m,n)$ as
\begin{equation}
\small
    \begin{aligned}
    &I_G^{Bin}(m,n) = \\ 
    &\begin{cases}
        \frac{I(2m-4:2:2m+2,2n-1:2:2n+1)+I(2m-1:2:2m+1,2n-4:2:2n+2)}{16} &\text{case 1} \\
        \frac{I(2m-1:2:2m+1,2n-2:2:2n)}{4}                              &\text{case 2} \\
        \frac{I(2m-2:2:2m,2n-1:2:2n+1)}{4}                              &\text{case 3}\\
        \frac{I(2m-3:2:2m+3,2n-2:2:2n)+I(2m-2:2:2m,2n-3:2:2n+3)}{16}    &\text{case 4}.\\   
    \end{cases}
    \end{aligned}
    \label{eqn:binning:green:final}
\end{equation}

\subsubsection{Sensor-Pixel Correspondences for Line-Skipping}
\label{apdx:line}
Line-skipping can be implemented in multiple ways. In this paper, we assume that line-skipping is implemented by removing 
every $2 \times l + 1$ and $2 \times l + 2$ (where $l$ is a natural number) rows and columns from the sensor output (as shown in Fig.~\ref{Fig:resizing schemes}). In other words, every $3^{rd}$ and $4^{\rm th}$ rows and columns are skipped. 

% Suppose $I^{temp}_R(m,n)$ denotes the red color component of the $(m ,n)^{\rm th}$ pixel the line-skipped sensor. Then 
% %When two rows and two columns (every $3rd$ and $4th$ rows/columns) are skipped for line-skipping (as in Fig.~\ref{Fig:lineskip}), red component goes through the following process:
% \begin{equation}
%     I^{temp}_R(m,n) = 
%     \begin{cases}
%         I(2m-1,2n-1)  &~~~~~~~~~~\text{case 1}\\
%         0             &~~~~~~~~~~\text{otherwise}.
%     \end{cases}
%     \label{eq:line:red}
% \end{equation}

% After demosaicing, $I_R^{Line}(m,n)$, denoting the red component of the $(m, n)^{\rm th}$ pixel index for the line-skipped image, can be found as
% \begin{equation}
%     I_R^{Line}(m,n) = 
%     \begin{dcases}
%         I^{temp}_R(m,n)                          &~\text{case 1}  \\
%         \frac{I^{temp}_R(m,n:2:n+2)}{2}          &~\text{case 2} \\
%         \frac{I^{temp}_R(m:2:m+2,n)}{2}          &~\text{case 3}\\
%         \frac{I^{temp}_R(m:2:m+2,n:2:n+2)}{4}    &~\text{case 4}.\\   
%     \end{dcases}
%     \label{eq:line:red1}
% \end{equation}
% Putting~(\ref{eq:line:red}) in~(\ref{eq:line:red1}), we obtain
% \begin{equation}
%     I_R^{Line}(m,n) = 
%     \begin{dcases}
%         I(2m-1,2n-1)                             &\text{case 1}  \\ 
%         \frac{I(2m-1,2n-3:4:2n+1)}{2}            &\text{case 2} \\ 
%         \frac{I(2m-3:4:2m+1,2n-1)}{2}            &\text{case 3}\\ 
%         \frac{I(2m-3:4:2m+1,2n-3:2:2n+1)}{4}     &\text{case 4}.\\    
%     \end{dcases}
%     \label{eqn:line:red}
% \end{equation}

Suppose $I_G^{temp}(m,n)$ denotes the green color of the $(m, n)^{\rm th}$ pixel of the line-skipped image. Then $I_G^{temp}(m,n)$ found as
\begin{equation}
\small
\begin{aligned}
&    I_G^{temp}(m,n) = 
    \begin{cases}
        I(2m-1,2n)         &~~~~~~~~~~\text{case 2} \\
        I(2m,2n-1)         &~~~~~~~~~~\text{case 3} \\
        0                  &~~~~~~~~~~\text{otherwise.}
    \end{cases}
\end{aligned}
\end{equation}

% ---- HERE WE NEED ONE MORE MIDDLE EQUATION, I THINK ----

Now, the green color of the $(m, n)^{\rm th}$ pixel of the line-skipped and then demosaiced image, $I_G^{Line}(m,n)$, can be obtained as
\begin{equation}
% \small
\begin{aligned}
    & I_G^{Line}(m,n) = \\ 
    &\begin{cases}
        \frac{I(2m-4:4:2m,2n-2)+I(2m-2,2n-4:4:2n)}{4}      &\text{case 1} \\ 
        {\scriptstyle I(2m-1,2n) }              &\text{case 2} \\
        {\scriptstyle I(2m,2n-1) }               &\text{case 3} \\
        \frac{I(2m-3:4:2m+1,2n-2)+I(2m-2,2n-3:4:2n+1)}{4}  &\text{case 4}.\\   
    \end{cases}
    \label{eqn:line:green}
\end{aligned}
\end{equation}

{\color{black}
\subsubsection{RoA of binning and bilinear scaling for increasing $k$}
An important issue is to determine how the RoA changes as the resizing factor changes. Consider, one of the images (i.e., video frame) being resized via binning of $k\times k$ and the other with bilinear scaling with a factor $\frac{1}{k}$ where $k \geq 2$. Here, we show how  RoA is affected by increasing $k$. 

In the binning scheme (as in Fig.~\ref{Fig:binning}), each $2\times2$ block is obtained from $4\times k^2$ pixels for red or blue pixels, and $6\times k^2$ pixels for green pixels from the raw image ($I^{raw}$). On the other hand, the number of pixels for bilinear scaling depends on $k$. When $k=2$, bilinear scaling will always use $9$ red, $12$ green pixels; when $k=3$, $12$ red and $16$ green; and when $k\geq4$ and $16$ red, $24$ green pixels from $I^{raw}$. As has been observed, while the number of pixels used in binning is increasing exponentially, the increase is lower or no increase for bilinear scaling. Table~\ref{tab:roa:calculated} shows the RoA for increasing values of $k$ obtained using the same approach as above.% which we estimated similar to the above sections.

\vspace{-2mm}
\captionsetup{labelfont={color=black},font={color=black}}
\begin{table}[h]
\centering
\caption {RoA for increasing $k$.}
\vspace{-1mm}
{\color{black}\begin{tabular}{|c|c|c|c|}
\hline
k & 2  & 3 & 4   \\ \hline
maximum & 0.53  & 0.31  &  0.25  \\ \hline
actual  & 0.46  & 0.22  & 0.14  \\ \hline
\end{tabular}}
\label{tab:roa:calculated}
\end{table}

\captionsetup{labelfont={color=black},font={color=black}}
\vspace{-1mm}

The table shows that as $k$ increases, the RoA between bilinear scaling and binning decreases. This indicates when a camera uses binning of $k\times k$ for $k\geq3$ for a video, using bilinear scaling for image is less likely to match.
Another issue is we didn't consider vertical and horizontal binning (i.e., the number of binned pixels are different along x- and y-axis, e.g., $3\times2$ is a vertical binning). Because these schemes may involve line-skipping, they can have various possible implementations. 
% Although we haven't seen any analysis or research on binning of $5\times5$ or higher, and higher values of $k$ will remarkably suffer blurring..
}

}

\newcommand{\ntk}{${\color{black}\forall}$}
\newcommand{\nntk}{${\color{black}\exists}$}

\vspace{-7mm}
\section{Analysis for Cameras}
\label{appendix:cameras}
This appendix is a brief documentation of the characteristics of the cameras contained in the NYUAD mixed media dataset as revealed by our study. As explained in Section~\ref{sec:approach}, the dataset contains $62$ models from $19$ camera brands.

The $1^{st}-3^{rd}$ columns in the table are the brand, model, number of devices, respectively. %For example, the first camera's brand is Asus and model is Z00AD. Visual media from only one camera is present from this model in the dataset. 
The $4^{\rm th}$ column shows if a model uses boundary pixels for capturing videos. $12$ of $62$ models use boundary pixels for at least one video resolution. So failing to address this issue will result in a mismatch decision when images and videos of those cameras are correlated. Specifically, all Lenovo cameras, Nexus $6$, Micromax E311 and so on use boundary pixels for capturing videos. %Since \cite{taspinar2016source} had failed to address this issue, their match ratio was relatively lower. 

The $5^{\rm th}$ and $6^{\rm th}$ columns show the cameras whose images and videos can match ``only by resizing" (cropping ratio is $1$) or require an exhaustive search. The importance of these two columns is they show cameras whose images can be quickly matched with their videos. This depends on the fact that if videos are captured with active image (with or without boundary pixels), they can be matched after a few correlations.% For example, the images of almost all Samsung models can be attributed to their source cameras using ``only by resizing" whereas the images from none of the Apple models require an exhaustive search. 

% The $7^{\rm th}$ and $8^{\rm th}$ columns indicate the same  for correlation of two different resolution videos from the same camera. As seen, videos of same source in the dataset can be matched with each other by aspect resizing except $3$ cameras: Samsung Galaxy A7, Google Nexus 4, and Yuraka AO5510. Videos of different resolution from these cameras require to test all possible resizing factors. 

{\color{black}The last two columns show minimum and maximum cropping ratios of videos in the dataset. This shows majority of cameras captured videos with a small cropping ratio which indicates a match can occur in a short period of time for them. N/A ones are due to the fact that we were not able to match the videos of those cameras to their images.}

The information provided here can be used for speeding up source attribution in multiple  ways. For example, Motorola G3 devices capture videos using boundary pixels and can be matched with images by only resizing. Given this information, we can, first crop the boundary pixels and match the videos with images by only resizing. Without this information, one may need to exhaustively try all possible resizing factors.

% Another example is when we correlate two videos, we can match them using aspect resizing for all the cameras in our dataset except when the camera is Samsung Galaxy A7, Google Nexus 4, and Yuraka AO5510. Unless we know the videos are captured by one of these three models, we can simply do aspect resizing, and quickly do the verification process. 

Note that in the table, $\forall$ indicates the condition is true for all cases whereas $\exists$ only for some cases. %As an example, suppose we have two videos of different resolution from two different cameras that are HTC Desire $526$. To make sure, they are not from the same camera, we should do exhaustive search on resizing factor as none of the scaling will yield a match decision. But using this information, we can simply do $2-$time correlation (via aspect resizing) and give the decision. Same applies to mixed media setting. 

%To sum up, this brief table contains some useful information to help forensics researchers and analysts. A more detailed camera having the exact scaling and cropping parameters of mixed media from a more significant number of media can be more useful.
%The table can be expanded to all available cameras such that a more clear schema for cameras can be more clearly seen.

%For example, all Lenovo cameras in our dataset uses boundary pixels for videos. 

\setlength{\topmargin}{-0.5in}
\newcommand{\rbm}{\rotatebox[origin=c]{90}} 
\begin{table*}[ht]
\small
\centering 
\label{append:a} 
 
\begin{tabular}{|c|c|c|c|c|c|>{\color{black}}c|>{\color{black}}c|} 
\hline 
\multicolumn{3}{|c|}{}                                    & ~~ video capture ~~  &  \multicolumn{2}{c|}{~~image vs video~~}  &  \multicolumn{2}{c|}{{\color{black}crop factor}}\\ \hline

Brand                      &  Model             &  \rotatebox[origin=c]{90}{ \#device }  &   \rbm{ Boundary } \rbm{ pixels }  & \rbm{ resizing }  \rbm{ only } &  \rbm{ exhaustive }  &  \rbm{min}   &  \rbm{max}  \\ \hline
\multirow{2}{*}{Asus}      &  Z00AD             &1&    &   &  \ntk   &1.3& 1.313\\
                           &  Z00LD             &1&    &  \ntk  &    &~~0.998~~~& ~~0.999~~~\\ \hline
Coolpad                    &  Note 3            &1&    &  \ntk  &    &0.998& 0.999\\ \hline
\multirow{17}{*}{Samsung Galaxy}  &  A3         &1&    &  \ntk  &  &0.999& 0.999\\
                           &  A5         &1&    &  \ntk  &  &0.998& 0.999\\
                           &  A7         &1&    &  \nntk  &  \nntk  &0.998& 1.339\\ 
                           &  A8         &1&    &  \ntk  &  &1.002& 1.002\\
                           &   E7         &1&    & \nntk & \nntk &1.4& 1.400\\
                           &   J5         &2&    &  \ntk  &  &0.998& 0.999\\
                           &   J7         &1&    &  \ntk  &  &0.998& 0.999\\
                           &   S4         &1&    &  \ntk  &  &0.998& 1.000\\
                           &   S6 Edge    &2&    &  \ntk  &  &0.994& 1.000\\ 
                           &   S7         &1&    &  \ntk  &  &0.998& 0.999\\
                           &   Note 1  &1&    &   &  \ntk  &1.572& 1.572\\
                           &   Note 2     &1&    &  \ntk  &  &1.002& 1.002\\
                           &   Note 3     &1&    &  \ntk  &  &1& 1.002\\
                           &   Ace 3      &1&    &  \ntk  &  &0.998& 0.998\\
                           &   S Duos     &1&    &  \ntk  &  &1& 1.000\\
                           &  Grand Prime       &1&    &  \ntk  &  &0.998& 0.999\\
                           &   Grand 2    &1&    &  \ntk  &    &0.992& 1.001\\ \hline
\multirow{4}{*}{Google}    &  Nexus 5           &3&    &  \nntk  &  \nntk &0.981& 0.999\\
                           &  Nexus 4           &1&    &  \ntk &  &0.993& 1.020\\
                           &  Nexus 6           &1&  \ntk  & \ntk &  &0.991& 0.991\\
                           &  Nexus 6p          &1&    &  & \ntk &1.252& 1.252\\ \hline
\multirow{6}{*}{HTC}       &  Desire 526        &1&    &  \ntk  &  &0.996& 1.004\\
                           &  Desire 620G       &1&    &  \ntk  &  &0.999& 1.000\\
                           &  Desire 626G       &1&    &  \ntk  &  &1.001& 1.006\\
                           &  Desire 628        &1&    &  \ntk  &  &0.998& 1.000\\
                           &  Desire 820s       &1&    &  \ntk  &  &0.998& 1.001\\
                           &  Desire 826        &1&    &  & \ntk &1.099& 1.101\\ \hline
\multirow{3}{*}{Apple}     &  iPhone 5          &1&    &  & \ntk &1.168& 1.560\\
                           &  iPhone 5s         &1&    &  &  & N/A & N/A\\
                           &  iPhone 6          &1&    &  & \ntk &1.2& 1.200\\ \hline
\multirow{4}{*}{Lenovo}    &  K3 Note           &3&  \ntk  & \nntk & \nntk &0.986& 1.318\\
                           &  Vibe K5           &1&  \ntk  & \ntk &  &0.983& 0.983\\
                           &  S90               &1&  \ntk  & \ntk &  &0.973& 0.975\\
                           &  P1a42             &1&  \ntk  & \ntk &  &0.973& 0.975\\ \hline
\multirow{2}{*}{Le X}      &509&1&  \ntk  &  \ntk &  &0.988& 1.001\\
                           &526&1&    &  & \ntk &1.08& 1.094\\ \hline
Intex                      &  Aqua Power        &1&    &  \ntk  &  &0.998& 1.001\\ \hline
Microsoft                  &  Lumia 640 XL      &1&    & \nntk & \nntk &0.994& 1.002\\ \hline
\multirow{3}{*}{Micromax}  &  A107              &1&    &  \ntk  &  &0.996& 0.997\\
                           &  E311              &1&  \ntk  &  \ntk &  &0.987& 1.009\\ 
                           &  Unite 3 Q372      &1&    &  \ntk  &  &0.998& 0.998\\ \hline
\multirow{5}{*}{Motorola}  &  G3                &2&  \ntk  &  \ntk  &  &0.955& 0.999\\
                           &  G4                &1&    &  \ntk  &  &0.998& 0.999\\
                           &  G4 plus           &1&    &  \ntk  &  &0.998& 0.999\\
                           &  G5 plus           &2&    &  \ntk  &  &0.998& 0.999\\
                           &  Moto X            &1&    &  & \ntk &1.127& 1.127\\ \hline
\multirow{2}{*}{One Plus}  &  One               &2&    & \nntk & \nntk &0.998& 1.444\\
                           &  X                 &1&    & \nntk & \nntk &0.987& 1.319\\ \hline
\multirow{2}{*}{Oppo}      &  A57               &2&  \ntk  & \ntk &  &0.994& 0.994\\
                           &1201&1&    &  &  & N/A & N/A\\ \hline
\multirow{3}{*}{Xiaomi}    &  Redmi 4           &1&    &  &  & N/A & N/A\\
                           &  Redmi Note 3      &5&  \ntk  & \ntk &  &0.987& 0.995\\
                           &  Redmi Note 4      &3&  \ntk  & \ntk &  &0.987& 0.989\\ \hline
\multirow{2}{*}{Sony}      &  Xperia M4         &1&    &  \ntk  &  &0.999& 0.999\\
                           &  Xperia C3         &1&    &  \ntk  &  &0.998&0.998\\ \hline
Vivo                       &  V5                &1&    &  \ntk&  &0.991& 0.991\\ \hline
Xolo Black                 &  1X                &1&  \ntk  & \ntk &  &0.986& 0.987\\ \hline
Yureka                     &  AO5510            &1&    & \nntk & \nntk &1.03& 1.072\\ \hline

\end{tabular} 
\end{table*}

\end{appendices}

\end{document}